\documentclass[a4paper,11pt]{article}
\pdfoutput=1 % if your are submitting a pdflatex (i.e. if you have
\usepackage{jcappub} % for details on the use of the package, please
\usepackage{cases}
\usepackage{amsmath}
\usepackage{bm}
\usepackage{braket}
\usepackage{latexsym}
\usepackage{graphicx}
\usepackage{subfigure}
\usepackage{hyperref}
\usepackage{amssymb}
\usepackage{float}

\newcommand{\be}{\begin{equation}}
\newcommand{\ee}{\end{equation}}
\newcommand{\ba}{\begin{eqnarray}}
\newcommand{\ea}{\end{eqnarray}}

\def\nue{\mathrel{{\nu_e}}}
\def\numu{\mathrel{{\nu_\mu}}}
\def\nutau{\mathrel{{\nu_\tau}}}

\def\barnu{\mathrel{{\bar \nu}}}

\def \lta {\mathrel{\vcenter{\hbox{$<$}\nointerlineskip\hbox{$\sim$}}}}
\def \gta {\mathrel{\vcenter{\hbox{$>$}\nointerlineskip\hbox{$\sim$}}}}

\def\t13{\mathrel{{\theta_{13}}}}
\def\y12{\mathrel{{\tan^2 \theta_{12}}}}
\def\c2{\mathrel{{\chi^2 }}}

\newcommand{\n}{neutrino}
\newcommand{\ns}{neutrinos}
\newcommand{\bg}{background}

\newcommand{\abs}{absorption}

\newcommand{\cnb}{C$\nu$B}

\title{Ultra High Energy Neutrinos: Absorption, Thermal Effects and Signatures}
\author{Cecilia Lunardini,}
\author{Eray Sabancilar,}
\author{Lili Yang}
\affiliation{Physics Department, Arizona State University, Tempe, Arizona 85287, USA.}
\emailAdd{Cecilia.Lunardini@asu.edu}
\emailAdd{Eray.Sabancilar@asu.edu}
\emailAdd{lyang54@asu.edu}

\abstract{We study absorption of ultra high energy neutrinos by the cosmic neutrino background, with full inclusion of the effect of the thermal distribution of the background on the resonant annihilation channel.  For a hierarchical \n\ mass spectrum (with at least one \n\ with mass below $\sim 10^{-2}$ eV),  thermal effects are important for ultra high energy neutrino sources at $z \gta 16$. The \n\ transmission probability shows no more than two separate suppression dips since the two lightest mass eigenstates contribute as a single species when thermal effects are included. 
Results are applied to a number of models of ultra high energy \n\ emission.  Suppression effects are strong for sources that extend beyond $z \sim 10$, which can be realized for certain top down scenarios, such as superheavy dark matter decays, cosmic strings and cosmic necklaces. For these, a broad suppression valley should affect the \n\ spectrum at least in the energy interval $10^{12} - 10^{13}$ GeV --   which therefore is disfavored for ultra high energy \n\ searches -- with only a mild dependence on the \n\ mass spectrum and hierarchy.  The observation of \abs\ effects would indicate a population of sources beyond $z \sim 10$, and favor top-down mechanisms; it would also be an interesting probe of the physics of the relic \n\ background in the unexplored redshift interval  $z \sim 10 -100$. 
}

\begin{document}
\maketitle
\flushbottom

%========================================================================================
\section{Introduction}
\label{sec:intro}

It is remarkable how our knowledge of the \n\ sky has so far been limited to a narrow band, from keV to a few TeV of energy, only recently reaching the PeV scale \cite{Aartsen:2013bka}.  Extending this range, in both directions, will reveal new phenomena, some of completely unknown nature so far.  The high energy frontier is expected to be especially rich, with \ns\ from baryonic accelerators (gamma ray bursts, active galactic nuclei, etc.), extending up to about $10^9$ GeV or so.  At even higher energy, the ultra high energy  (UHE) regime  will show cosmogenic neutrinos, and may even reveal the existence of  topological defects, which could emit \ns\ through a number of energy loss channels.   For those sources far beyond the gamma ray horizon (redshift $z\sim 0.1$ for 1 TeV energy \cite{gilmore}) \ns\ might very well be the only probe, since, even at the highest energies,  they propagate freely up to cosmological distances \cite{Berezinsky:1991aa}. 

At this time, experiments dedicated to the UHE regime are advancing rapidly: after a first successful phase, a new generation of experiments is expected to come online, and start to probe the parameter space predicted by theory. 
The prospects of a detection of UHE \ns\ is of great motivation to model the expected fluxes in detail, including a number of propagation effects like flavor oscillations and absorption, and possible other phenomena due to still unknown neutrino properties, like \n\ decay or non-standard interactions.  Neutrino absorption is the focus of the present work. 

Absorption of UHE \ns\  is largely due to scattering on the cosmic \n\ background (\cnb), and, for energies $E \gtrsim 10^{11}$ GeV, it establishes a \n\ ``horizon'' of $z \sim 140$, beyond which the Universe is opaque to \ns\ \cite{Berezinsky:1991aa}. In fact, the shape of this horizon is rather complicated, because the \n\ mean free path is resonantly suppressed -- depending on the \n\ energy and mass -- due to  \n-antineutrino annihilation via the $Z^0$ boson ($\nu + \barnu \rightarrow {\rm anything} $). The signature of this process is one or more characteristic absorption dips in the \n\ spectrum.

The phenomenology of  resonant \abs\  has been studied in various contexts, starting with the so called ``Z-dip" scenario \cite{Weiler:1982qy,Weiler:1983xx}, and explored further \cite{Weiler:1997sh,Fargion:1997ft} to explain possible cosmic ray events beyond the Greisen Zatsepin Kuzmin (GZK) cutoff of cosmic rays \cite{Greisen:1966jv,Zatsepin:1966jv}.  Subsequent works focused on modeling the sharp dips expected for non-relativistic \n\ background \cite{Roulet:1992pz,Eberle:2004ua}. It was then considered that, at least for the lowest of the three \n\ masses, thermal effects on the \cnb\ might be important. These were studied  in detail for a single \n\ species \cite{D'Olivo:2005uh}, and the transmission probabilities were calculated in the context of a more realistic three-neutrino mass spectrum \cite{Barenboim:2004di}.

In consideration of the ongoing and planned experimental advances, it is timely to continue the study of this topic in more detail, with a stronger attention to applications to various specific models of UHE neutrino sources. To include all suppression effects is especially important, to estimate the level of sensitivity required for detection, and to  correctly interpret future observations. Considering our improving knowledge of the \n\ mixing matrix, and the expectation that the \n\ masses may be measured soon in the laboratory, it is likely that absorption signatures in UHE \n\ data will  be used effectively as probes of the physics of the sources and of the \cnb. Specifically, a strong (weak) suppression would favor a population of sources at distance comparable to (smaller than) the \n\ horizon.  The degree of suppression could also provide a direct detection of the \cnb, and probe, at least in principle, its density, momentum distribution and possible exotic properties.

\begin{figure}[h]
   \centering \includegraphics[width=0.67\textwidth]{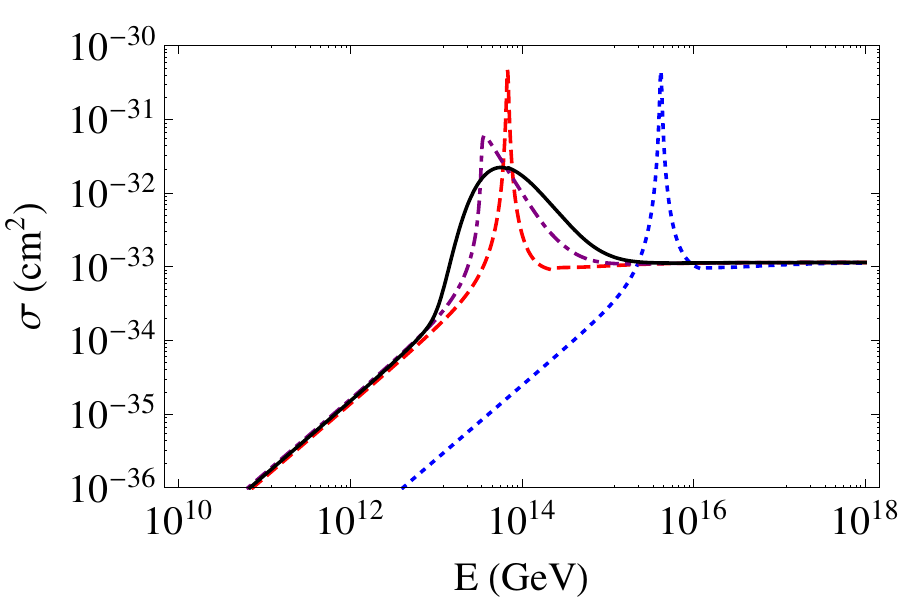}
 \caption{The total ($\nu + \bar \nu \rightarrow {\rm anything}$) cross section, as a function of the beam energy, for a thermal \n\ background of temperature $T = 1.714 \times 10^{-2}$ eV (corresponding to redshift $z=100$) and a background \n\ of mass $m= 10^{-3}$ eV. The curves correspond to  different levels of inclusion of thermal effects: (i) Dotted (blue): no thermal effects (target \n\ at rest), (ii) Dashed (red): same as (i), but with an effective value of the center of mass energy [see Eq.~(\ref{bars})], (iii) Dot-dashed (purple): for background \n\ momentum fixed at its root mean square value and averaged over the scattering angle, (iv) Solid (black):  full calculation (this work), after averaging over the background momentum distribution and scattering angle (see Sec.~\ref{sec:prop}). 
 }
\label{xsectionappr}
\end{figure}

Here we develop a fully realistic treatment of the absorption: considering the three active \n\ species, we include thermal effects exactly (Fig.~\ref{xsectionappr}), and discuss their dependence on the  \n\ mass spectrum. The results are applied to a number of proposed mechanisms of production of UHE \ns, including astrophysical sources, massive relics and topological defects.  The goal is to identify the main signatures at near future experiments, and discuss what physics can be learned from them. 

The paper is organized as follows. In Sec.~\ref{gen}, we give generalities on the detectability of UHE \ns, and discuss the physics of \ns\ and of the \cnb.  In Sec.~\ref{sec:prop}, absorption effects are discussed, with emphasis on thermal effects.  In Sec.~\ref{sec:uhen}, we consider a number of UHE \n\ sources, and for each of them, we discuss the signatures of \n\  \abs\ on the expected signal at Earth.    The results are then discussed in Sec.~\ref{sec:disc}.  Appendix \ref{app:cs} summarizes the  details of the calculation of the relevant \n\ cross sections (resonant and non-resonant).  In Appendix \ref{app:amplitude}, we elaborate on the effect of \n\ mixing on the scattering rate.

%===========Thermal Effect vs no-thermal effect

%========================================================================================
\section{Generalities}
\label{gen}

%===============================================
\subsection{Searching for UHE neutrinos: limits and sensitivity}

UHE neutrinos of energies $E \gtrsim 10^{10}$ GeV are very interesting as a testing ground for several high energy astrophysical mechanisms and top down models, as  well as of  new neutrino detection methods. The space mission JEM-EUSO \cite{Ebisuzaki:2008zza,Takahashi:2009zzc,MedinaTanco:2009tp} will use Earth's atmosphere as the target medium to detect fluorescent light from extensive air showers, and will be most sensitive at energies $E \sim 10^{10}-10^{11}$ GeV. Another efficient UHE neutrino detection method is the so called radio Cherenkov technique that is based on the Askaryan effect \cite{askaryan}. This effect was successfully tested and observed in a lab experiment \cite{Saltzberg:2000bk} and in Antarctic ice \cite{Gorham:2006fy}. Specific implementations of this technique have used the lunar regolith as target medium (GLUE \cite{Gorham:2003da}, NuMoon \cite{Scholten:2009zz,Buitink:2008bc,Buitink:2010qn} and RESUN \cite{Jaeger:2010zz}) or the ice of Greenland (FORTE \cite{Lehtinen:2003xv}) or the polar cap in Antarctica (ANITA \cite{Barwick:2005hn,Gorham:2008yk,Mottram:2012zz} and RICE \cite{Kravchenko:2006qc}).  A new generation of ongoing or planned initiatives will bring this method to its full potential. The main projects are LOFAR \cite{Scholten:2005pp,Scholten:2011zza,Buitink:2013gk}, SKA \cite{James:2008ff,Ekers:2009zz}, LUNASKA \cite{James:2009sf,Bray:2013ta}, ARIANNA \cite{Gerhardt:2010js}, AURA \cite{Landsman:2007zza,Landsman:2008rx} and EVA \cite{Gorham:2011mt}.

\begin{figure}[h]
  \centering \includegraphics[width=0.85\textwidth]{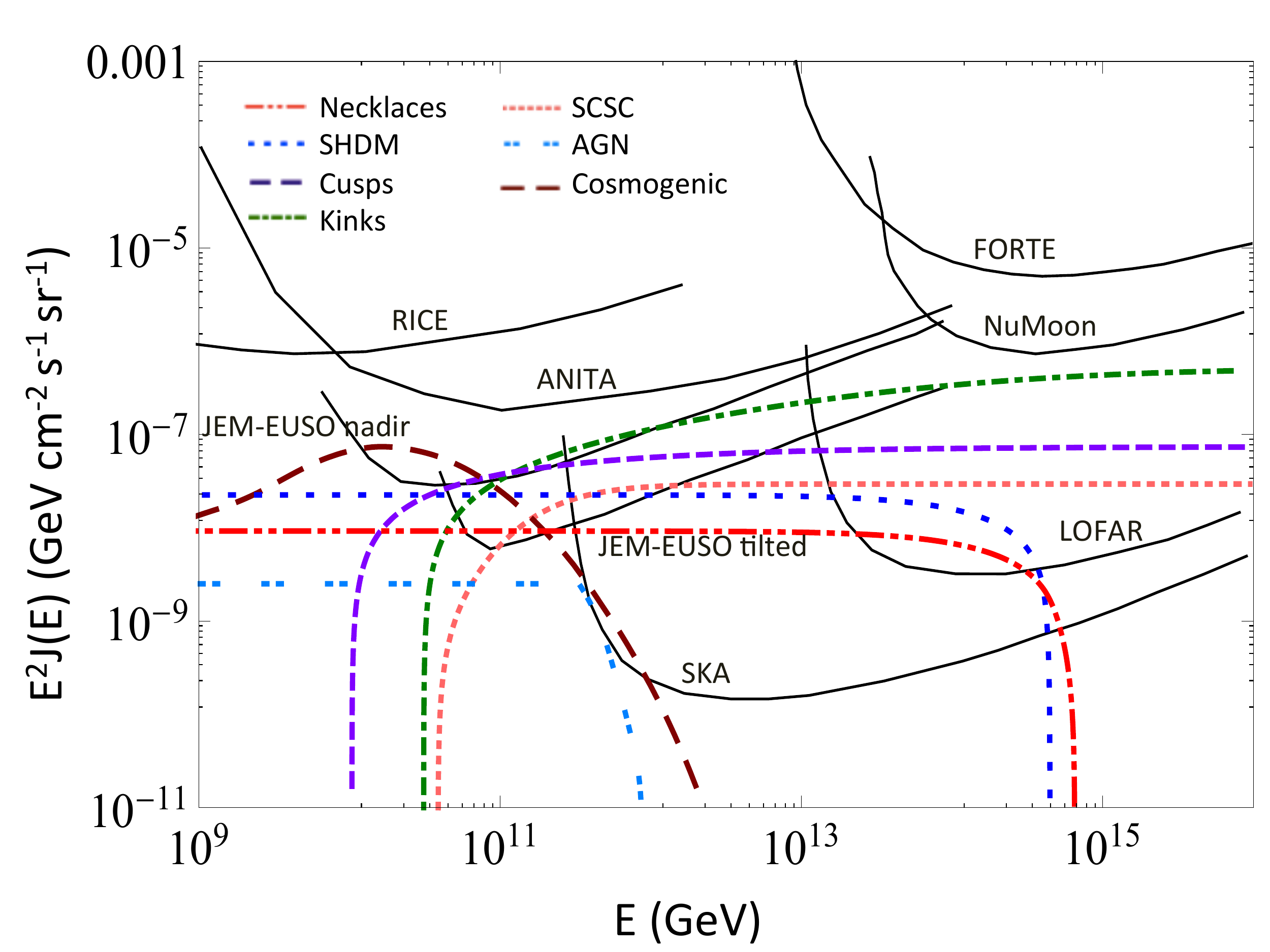}
 \caption{Solid (black) curves: existing upper bounds on the UHE neutrino flux from RICE, ANITA, FORTE, NuMoon, and expected sensitivities at JEM-EUSO (nadir and tilted modes), LOFAR and SKA.  Non-solid (color) curves:  UHE \n\ fluxes from cosmic string cusps, cosmic string kinks, superconducting cosmic string cusps (SCSC), cosmic necklaces, superheavy dark matter (SHDM), cosmogenic neutrinos and active galactic nuclei (AGN) (see the legend in the figure).}
\label{fluxeslimits}
\end{figure}

In Fig.~\ref{fluxeslimits}, we show the expected sensitivities from LOFAR, SKA and JEM-EUSO as well as the upper bounds on the UHE neutrino flux from RICE, ANITA, FORTE and NuMoon. The figure also shows some expected UHE neutrino fluxes from various sources, such as cosmogenic neutrinos, active galactic nuclei, superheavy dark matter decays, cosmic string cusps and kinks, superconducting cosmic strings and cosmic necklaces. We shall elaborate in detail on these in Sec.~\ref{sec:uhen}.  
%====Flux Limits

All the UHE neutrinos fluxes we consider here originate from hadronic cascades that produce numerous pions, which eventually decay into \ns, electrons, positrons and photons. About half the initial energy density in pions goes to electromagnetic energy density including high energy $e^{\pm}$ and $\gamma$-rays. These particles have very short mean free paths since they interact electromagnetically with the background photons (CMB and extragalactic background light). The initial electromagnetic energy density quickly cascades down to lower energy photons leaving a diffuse flux of $\gamma$-ray photons behind. Therefore, one can obtain an upper bound on UHE neutrino flux based on the observed diffuse $\gamma$-ray background, as was first considered by Berezinsky and Smirnov \cite{Berezinsky:1975zz}. The most recent upper limit is based on the Fermi-LAT observations \cite{Abdo:2010nz} of the $\gamma$-rays \cite{Berezinsky:2010xa}. We checked that the example fluxes shown in Fig.~\ref{fluxeslimits} are consistent with this cascade upper bound.

%===============================
\subsection{Neutrino masses and mixing}
\label{massmix}

For decades after its discovery, the \n\ has remained a mysterious particle in its fundamental properties. The question if \ns\ are massive or not has remained open for decades, during which laboratory limits have narrowed down the \n\ mass  to a few eV \cite{Kraus:2004zw,Aseev:2011dq}. This scale has been recently surpassed (although in a parameter-dependent way) by cosmological probes yielding upper limits on the sum of the neutrino masses. At $95 \%$ CL, the main limits are $\Sigma m_\nu < 0.44$ eV by the WMAP 9-year data \cite{Hinshaw:2012aka} and $\Sigma m_\nu < 0.23$ eV by the recent Planck data \cite{Ade:2013zuv}.  

\begin{figure}[h]
\centering \includegraphics[width=0.48\textwidth]{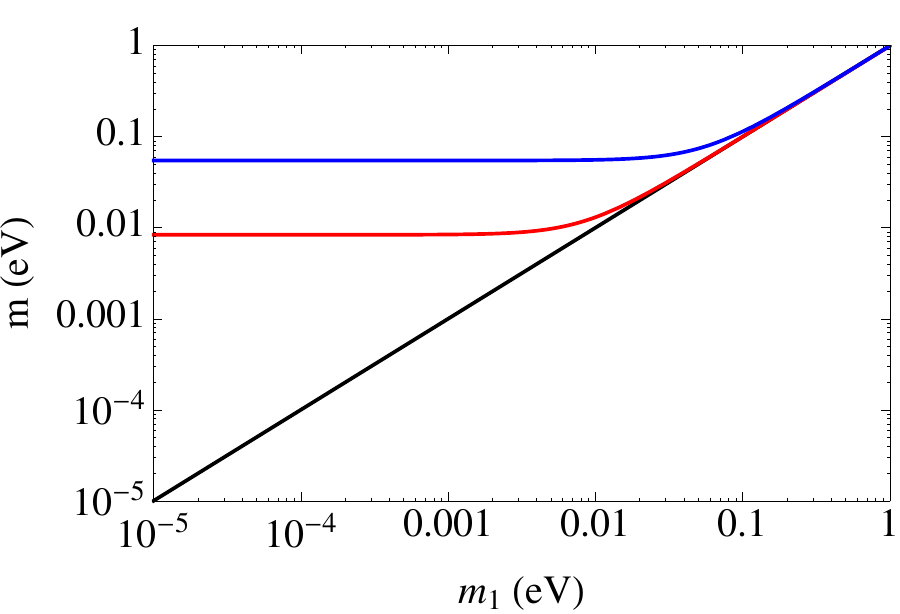}
\centering \includegraphics[width=0.48\textwidth]{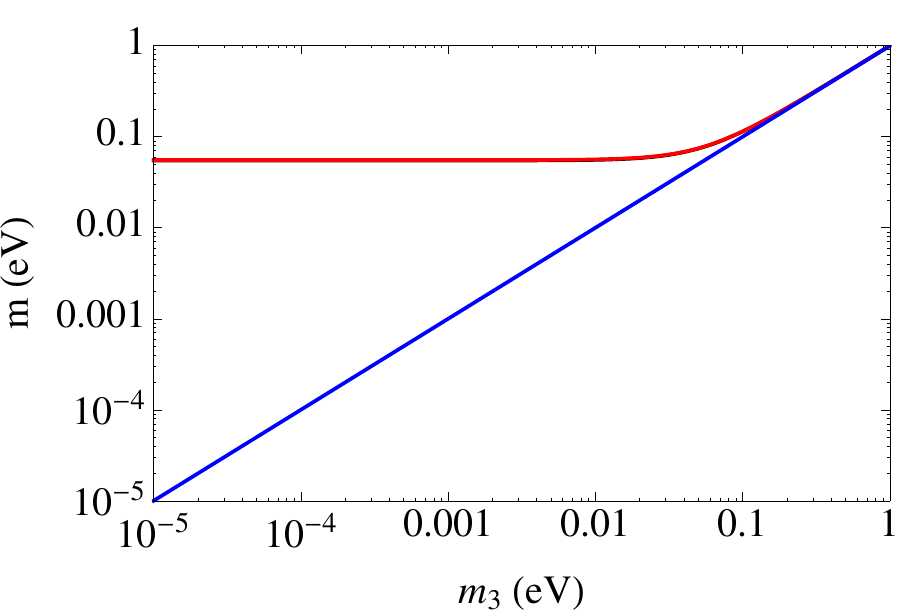}
\caption{The three  neutrino masses as a function of the minimum mass $m_{\rm min}$.    The left (right) panel  is for the normal (inverted) hierarchy, where $m_{\rm min}=m_1$ ($m_{\rm min}=m_3$). }
\label{nhmasses}
\end{figure}

The fact that \ns\ have mass has finally been established through the discovery of \n\ flavor oscillations. Indeed, in the absence of exotic \n\ interactions, oscillations are possible only if there are two physically distinct bases for a system of three \ns: the basis of the mass eigenstates, $\nu_i$ (with masses $m_i$, $i=1,2,3$), and the basis of eigenstates of the weak interaction (flavor states), $\nu_\alpha$ with $\alpha=e,\mu,\tau$.    Oscillation probabilities depend on the mixing matrix connecting the two bases, and on the mass squared differences $\Delta m^2_{ij}\equiv m^2_i - m^2_j$.   The latter are measured to be $\Delta m^2_{21}  \simeq 7.5 \times 10^{-5}~{\rm eV^2}$ and $| \Delta m^2_{31} | \simeq 2.4 \times 10^{-3}~{\rm eV^2}$ (see e.g., \cite{Fogli:2012ua}), indicating that:
\begin{enumerate}
\item There are two possibilities, or {\it hierarchies}: $m_1 <m_2 < m_3$ (normal hierarchy, NH) and $ m_3 <m_1 < m_2 $   (inverted hierarchy, IH).

\item At least two of the three masses are not zero: for NH, we have $m_2 \gta 8.6 \times 10^{-3}~{\rm eV}$ and $m_3 \gta 4.8 \times 10^{-2}~{\rm eV}$; for IH we get $m_1 \sim m_2 \gta 4.8 \times 10^{-2}~{\rm eV}$. 
Fig.~$\ref{nhmasses}$ illustrates the possible values of the three masses depending on the hierarchy.   We distinguish between a {\it hierarchical} mass spectrum, where at least two of the masses differ by one or more orders of magnitude, and a {\it  degenerate} spectrum with masses of comparable value. As the figure shows, the degenerate case requires the smallest mass ($m_{\rm min} =m_1$ or $m_3$ depending on the hierarchy) to exceed a few times $10^{-2}$ eV.
\end{enumerate}

%Mass hierarchy

Here we choose to work with eight  representative mass spectra (see Table~\ref{tabmass}), four for each hierarchy, where the smallest mass equals $m_{\rm min} = 10^{-5}, 10^{-3}, 2 \times10^{-2}, 8 \times10^{-2}$ eV. In order, these correspond to spectra that are extremely hierarchical, moderately hierarchical, moderately degenerate and very degenerate. 

%Masses used, for normal hierarchy (NH)
\begin{table}[h]
  \centering 
  \begin{tabular}{c|c|c}
   \hline 
    \hline 
$m_1$ &  $m_2$&  $m_3$ \\
\hline
 \hline 
 0.00001 &  0.0084 &  0.055 \\
 \hline
0.001 & 0.0084 &  0.055    \\
 \hline 
 0.02 & 0.022 & 0.058 \\
 \hline
0.08 &  0.080 & 0.097 \\ 
 \hline
  \hline
\end{tabular}
\qquad
 \begin{tabular}{c|c|c}
   \hline 
    \hline 
$m_1$ &  $m_2$&  $m_3$ \\
\hline
 \hline 
0.055 &  0.055 &  0.00001 \\
 \hline
 0.055 &  0.055 &  0.001  \\
 \hline 
0.058 &  0.059 &  0.02 \\
 \hline
 0.097 &  0.097 &  0.08 \\ 
 \hline
  \hline
\end{tabular}
  \caption{Values of the neutrino masses (in eV) used in this work for NH (left) and IH (right).}
  \label{tabmass}
\end{table} 

\begin{figure}[h]
\centering \includegraphics[width=0.6\textwidth]{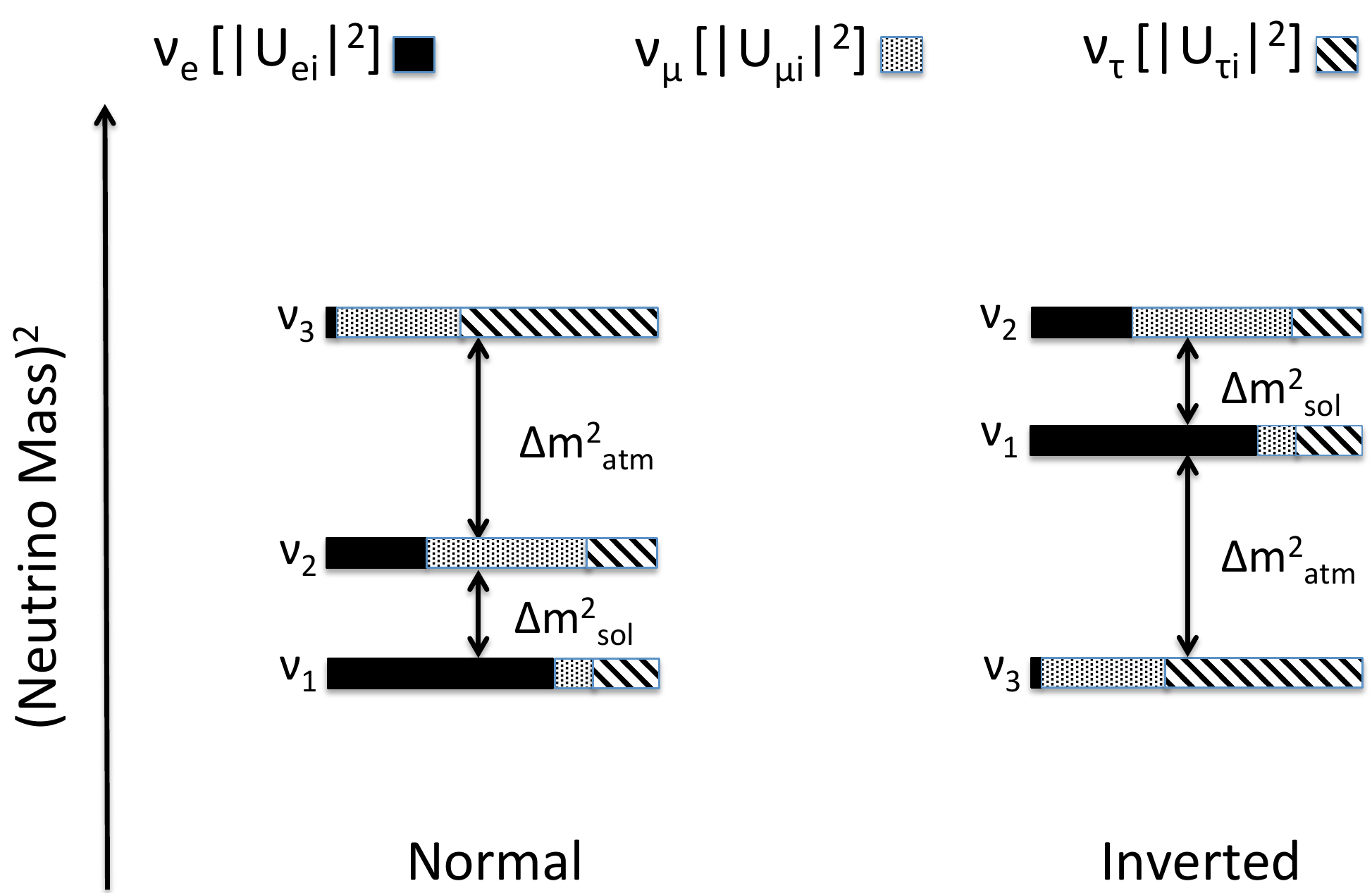}
\caption{A graphical illustration of the mixing between mass and flavor eigenstates. The boxes represent the mass eigenstates, $i=1, 2, 3$, the shaded regions represent their flavor admixtures $|U_{\alpha i}|^2$ for $\alpha = e, \mu, \tau$, $|\Delta m^2_{\rm atm}| = |\Delta m^2_{31}| \approx |\Delta m^2_{32}| = 2.4 \times 10^{-3}~\rm{eV}^2$ and $\Delta m^2_{\rm sol} = \Delta m^2_{21} = 7.5 \times 10^{-5}~\rm{eV}^2$. }
\label{boxdiagram}
\end{figure}
As will be seen, the mixing  of the mass eigenstates with each flavor state is important to establish the optical depth, and therefore the horizon, for \n\ of different flavors.   The structure of the PMNS mixing matrix \cite{Pontecorvo:1957cp,Pontecorvo:1967fh,Maki:1962mu} $U_{\alpha i}$ depends on three angles, $\theta_{ij}$, with  $\sin^2 \theta_{12}\simeq 0.31$,  $\sin^2 \theta_{23}\simeq 0.42$, and $\sin^2 \theta_{13}\simeq 0.025$ \cite{Fogli:2012ua}. The fact that $\theta_{23}$ is very close to $\pi/4$, while $\theta_{13}$ is relatively small, makes it so that $\nu_3$ is nearly a 50-50\%  mixture of $\numu$ and $\nutau$ with a small $\nue$ component, while $\nu_1$ and $\nu_2$ have large admixtures of all the three flavors (Fig.~\ref{boxdiagram}).

%==================================
\subsection{Cosmic neutrino background}
\label{sec:cnb}

We assume a Friedmann Roberston Walker (FRW), $\Lambda$CDM universe, with the Hubble parameter
\be
H(z) = \frac{ \dot a}{a}= H_{0} ~ \sqrt{\Omega_{\rm m} (1+z)^3 + \Omega_{\Lambda}}, 
\ee
where $H_{0} = 70.4$ Mpc/km/s, $a = 1/(1+z)$ is the scale factor, $z$ is the cosmological redshift, $\Omega_{\rm m} = 0.272$ and $\Omega_{\Lambda} = 0.728$  are the fraction of the energy density of matter and dark energy respectively \cite{Komatsu:2010fb}.  We also use natural units, $c= \hbar =1$, and set Boltzmann's constant $k=1$. The relationship between the proper time, $t$, and the redshift, $z$, is given by:
\be\label{dt}
dt = \frac{dz}{(1+z)H(z)},
\ee
and the comoving distance is
\be\label{dr}
dr = \frac{dz}{H(z)}~,
\ee
so that the comoving volume is given by
\be\label{dVc}
dV_{\rm c} = r^2 dr d\Omega~,
\ee
where $r=r(z)$ is the integral of Eq.~(\ref{dr}) from present epoch to redshift $z$. Thus, the physical volume is simply $dV(z) = dV_{c}/(1+z)^3$.

Standard cosmology predicts the relic abundance of neutrinos with a thermal spectrum, similar to the cosmic microwave background (CMB) photons. Thermal equilibrium is provided by weak interactions, hence the relic neutrinos are produced in flavor eigenstates. The number density of the \cnb\ for a single \n\ species, is given by the Fermi-Dirac distribution (assuming zero chemical potentials) at temperature $T$ as\footnote{After \n\ decoupling, $T$ is actually {\it defined} by Eq.~(\ref{dnnu}).} 
\be\label{dnnu}
dn(p,T) =  \frac{d^3 p}{(2\pi)^3} \frac{1}{e^{p/T} +1} ~ , 
\ee
with $p$ being the \n\ momentum.   Because both $p$ and $T$ scale as $(1+z)$, $dn$ simply scales as $d^3 p \propto (1+z)^3$.

A neutrino is produced  via the weak interaction as a flavor eigenstate, $\ket{\nu_{\alpha}}$. This state is a linear combination of mass (energy) eigenstates, $\{\ket{\nu_j}\}$, so that the time evolved freely streaming flavor ket has the form
\be
\ket{\nu_{\alpha} (t)} = \sum\limits_{j=1}^{3} U^\ast_{\alpha j}~ e^{- i\Phi_j(t)}~ \ket{\nu_j}~,
\ee
where $\Phi_j(t) = \int_{t_{i}}^{t} dt'~ \sqrt{[p(t')]^2 + m_{j}^2}$, and $t_{i}$ is the neutrino production time.
 As the \n\ propagates onward, its three mass eigenstate components drift along with different velocities (due to their different masses). Therefore, given enough time, they would eventually become spatially separated, so that the coherence of the \n\ wavepacket is lost. When and how decoherence occurs for the \cnb\ depends on the size of the \n\ wavepacket at decoupling, and on the \n\ mass spectrum; it has been studied only in part \cite{Fuller:2008nt,Bernardini:2012uf,Dodelson:2009ze}, with the general conclusion that, for a hierarchical mass spectrum, decoherence should have taken place between the \n\ horizon and today. 
Here we postpone a detailed description of \n\ decoherence, and consider the \cnb\ to be in mass eigenstates, in accordance with previous literature \cite{Eberle:2004ua}.  We expect that a more accurate treatment of the quantum state of the \cnb\ might perhaps change some quantitative details, but maintain the general features of the \abs\ pattern and the main conclusions of our work. 

Let us now estimate when thermal effects are important. A given mass eigenstate (or, better, most of its population) is non-relativistic at $z \lesssim z_{{\rm th}, j}$: 
\be\label{zd}
1+z_{{\rm th},j} \sim \frac{m_j}{\bar p_{0}} \sim 16 \left( \frac{m_j}{10^{-2}~{\rm eV}} \right)~. 
\ee   
where $\bar p_0 = \sqrt{< p^2>} = 3.597 \, T_{0} = 6.1044 \times 10^{-4}$ eV and $T_{0} = 1.697 \times 10^{-4}$ eV is the \cnb\ temperature at present epoch. 
Eq.~(\ref{zd}) means that the heaviest mass eigenstate ($m \gta 0.05~{\rm eV}$) is non-relativistic at $z \lesssim 83$.
 It is immediate to see that for $z \gta z_{{\rm th}, j}$ thermal effects should be substantial for the $\nu_j$ component of the \cnb\  in the scattering with UHE \ns. Therefore
 Eq.~(\ref{zd}) offers a good guidance of the range of production redshifts where these effects should be included.  The numerical results in the next section confirm this rough estimate.

%========================================================================================
\section{Neutrino absorption effects}
\label{sec:prop}

There are several propagation effects that leave imprints on the observed UHE neutrino fluxes. Since they are ultrarelativistic, their energy redshifts due to the expansion of the universe. Oscillations change the flavor composition of the \n\ flux, generally in the direction of  flavor democracy, where all flavors are equally represented.  Neutrinos are also absorbed due to resonant and non-resonant scatterings on the \cnb. As a result of all these processes, the observed \n\ flux can differ significantly from the one at the production point.  Here we focus on \abs, and incorporate the other effects as needed. The effect of cosmological redshift is included, and oscillations are modeled at the basic level  by assuming flavor equipartition. As it will be shown, this is sufficient because the results are not very sensitive to the flavor composition of the UHE flux.  

%========================
\subsection{Cross sections}
\label{sec:nuabs}

Several channels contribute to the process $\nu + \bar \nu \rightarrow {\rm anything}$ \cite{Roulet:1992pz,Barenboim:2004di}. Their contributions to the total cross section, $\sigma_{i}(E, p, m_j, z)$ (with $i$ indicating various channels), are summarized in Appendix  \ref{app:cs}, and shown in Fig.~\ref{xsectionpartial}.    

These cross sections depend on the physics of the colliding \ns\ through the  Mandelstam variable, 
\ba
&&s = (q^{\mu} + p^{\mu})^2 \approx 2 E' \left( \sqrt{p^2 + m_j^2} - p \cos\theta \right) , \label{sfull} \\
&&q^{\mu} = [E', {\bf  q}], \\
&&p^{\mu} = \left[ \sqrt{p^2 + m_j^2}, {\bf p} \right] ~,
\label{svar}
\ea
 where $q^{\mu} $   and  $p^{\mu} $ are the four momenta of the UHE \n\ (``beam neutrino" from here on) and the background \n\, respectively, and ${\bf  q} \cdot {\bf p} \equiv p~ q \cos\theta$. 
We use the approximation $E' = q$, since $q \gg m_j$ for the beam \n. 

There are two qualitatively different channels that contribute to the scattering of UHE neutrinos on the \cnb. The resonant channels correspond to annihilation of an UHE neutrino (antineutrino) with a background antineutrino (neutrino) via a $Z^0$-resonance in the s-channel. The resonance occurs at $\sqrt{s} = M_{Z}$, where $M_Z =91.1876$ GeV is the $Z^0$ boson mass, hence it is  sensitive to the value of the neutrino mass, $m_j$, as well as the background neutrino momentum, $p$. Although the details can be complicated (see Appendix \ref{app:cs}), the main dependence of the total  cross section on $s$, and therefore on $q^{\mu} $   and  $p^{\mu} $ shows three regimes (see Fig.~\ref{xsectionpartial}):
\begin{enumerate} 
\item {\it Sub-resonance}: the cross section depends linearly on the  energy of the beam \n, $E'$.
\item {\it At or near resonance}: the cross section is dominated by the resonant term, which has the characteristic Breit-Wigner enhancement.
\item {\it Above resonance}: the non-resonant cross sections approach an asymptotic value $\sigma_{\rm nr} \sim 10^{33}~{\rm cm^2}$. 
\end{enumerate}
%=Cross section, different channels
\begin{figure}[h]
  \centering \includegraphics[width=0.75\textwidth]{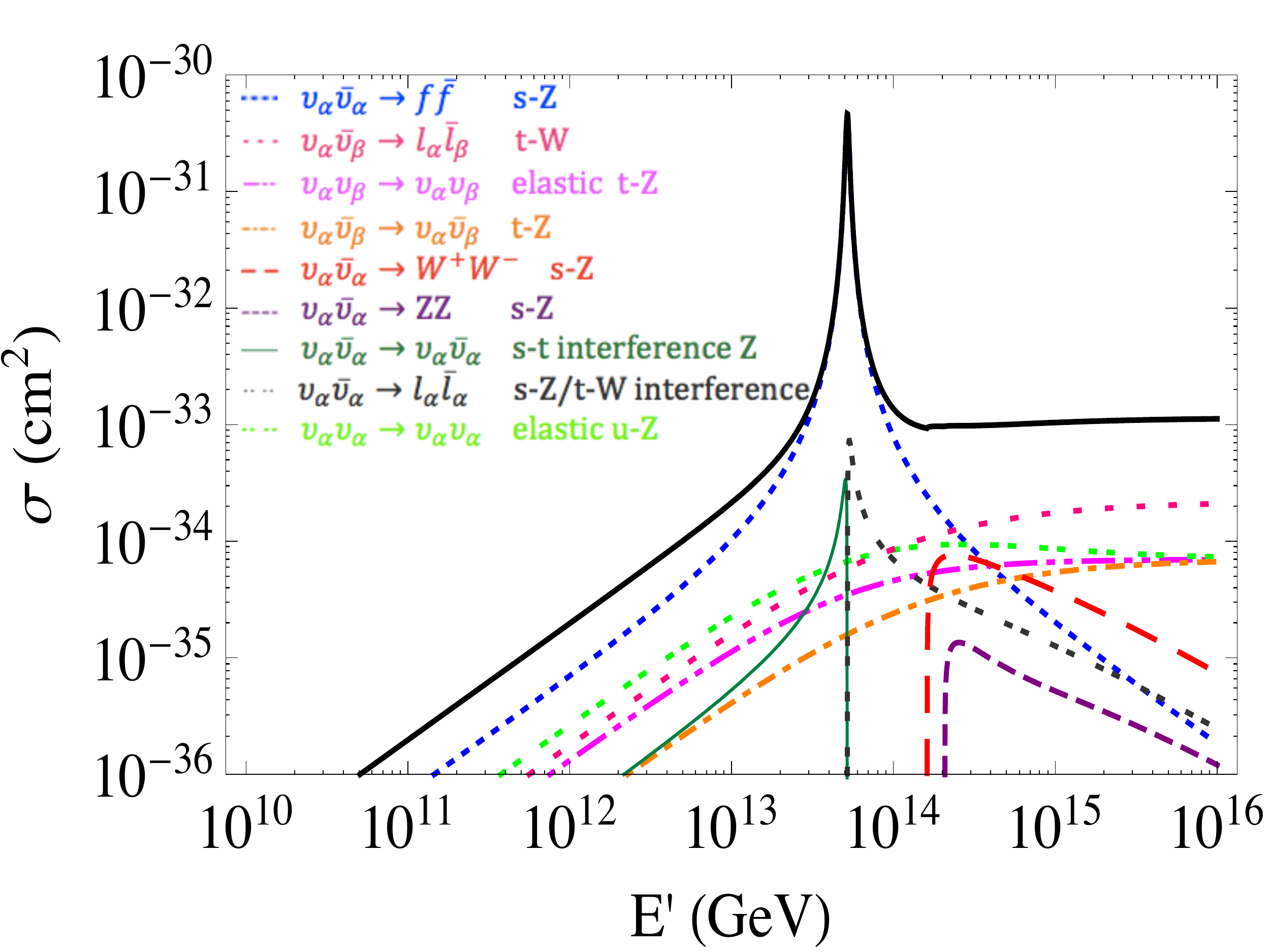}
 \caption{Contributions to the cross section for the processes $\nu + \bar \nu \rightarrow {\rm anything}$. Various channels are shown with thin (colored) lines (see the legend in the figure), and the total cross section is shown with a thick (black) line for a representative \n\ mass of $m_\nu =0.08$ eV. Thermal effects are not included.}
\label{xsectionpartial}
\end{figure}

In Fig.~\ref{xsectionpartial}, these regimes are shown for the simplest and the most discussed case of the background \n\ at rest, i.e., $m_j \gg  p$, which is realized for the \cnb\ at the present time and $m_j \gta  10^{-3} $ eV.  In this limit, $s \approx 2E' m_j  $, therefore the resonance energy,  $E_{\rm res}^\prime \approx M^2_{Z}/(2 m_j)$ is a clear identifier of the \n\ mass. 

The situation is less transparent in the most general case, where the momentum of the background  \ns\ is not negligible, and thus the exact form of $s$ given by Eq.~(\ref{sfull}) must be used.  This is the central purpose of our study: to work out the phenomenology of \n\ absorption in its generality, and apply it to realistic examples.  To do so, we have studied, for the resonant channels: \\

\begin{figure}[h]
  \centering \includegraphics[width=0.49\textwidth]{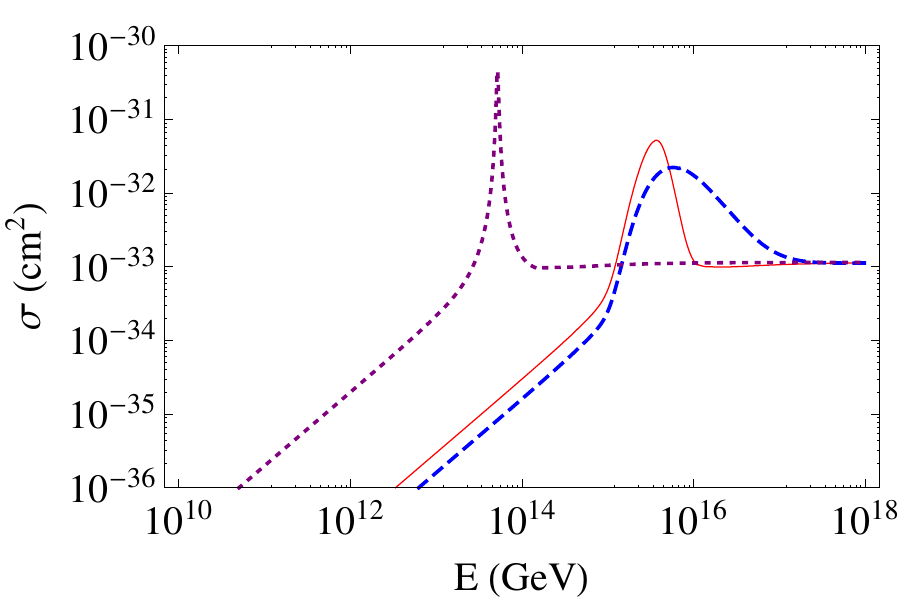}
   \centering \includegraphics[width=0.49\textwidth]{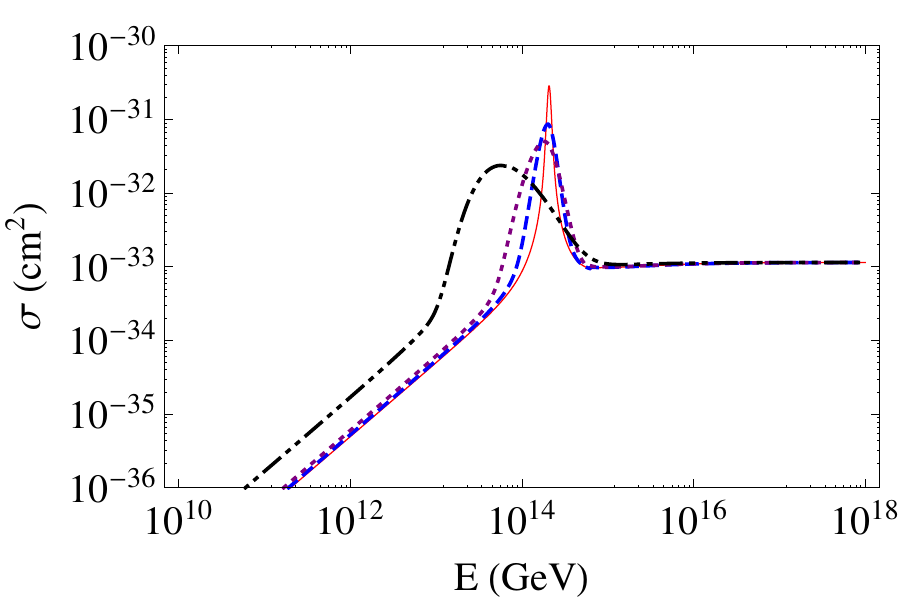}
  \caption{Left panel:  the total ($\nu + \bar \nu \rightarrow {\rm anything}$) cross section, inclusive of thermal effects, when averaged over a \n\ background with a momentum distribution as in Eq.~(\ref{dnnu}), with temperature $T_0=1.697 \cdot 10^{-4}$ eV.  Different lines correspond to \n\ masses: $m_\nu/ {\rm eV} =8 \times 10^{-2}~({\rm purple-dotted}),~10^{-3}~({\rm solid-red}),~10^{-5}~({\rm dashed-blue}) $. Right panel:   the same cross section, for the \n\ mass of $m_\nu =2 \times 10^{-2}$ eV and a \n\ background with temperatures $T/ {\rm eV}= 3.394 \times 10^{-4}, 1.867 \times 10^{-3}, 3.56 \times 10^{-3},1.713 \times 10^{-2}$ corresponding to $z= 1~({\rm red-solid}),~ 10~({\rm blue-dashed}),~ 20~({\rm purple-dotted}),~ 100~({\rm black-dash-dotted})$.  }
\label{xsectionthermal}
\end{figure}

\begin{enumerate}
\item {\it The differential cross section $d\sigma_Z/d\Omega$}.  
This depends on $\theta$ through $s$, given by  Eq.~(\ref{sfull}) (see also Appendix \ref{app:cs}).  Since the \cnb\ is isotropic,  we will only need the total cross section, $\sigma_Z$, obtained by integrating  $d\sigma_Z/d\Omega$ over the angular variables.  The result \cite{D'Olivo:2005uh} is given in Appendix \ref{app:cs}. Although its explicit expression is complicated, its main features can be understood considering that now the resonance is realized for an interval of the beam \n\ energy, corresponding to $\theta$ varying between $0$ and $\pi$ [see Eq.~(\ref{sfull})]. This results in a spread in the resonance peak compared to the background at rest (see Fig.~\ref{xsectionappr}).  The cross section at resonance is larger for a head-on collision, $\theta = \pi$. This is because, there, the energy $E'$ required to realize the resonance is minimum, and therefore the prefactor $1/E'$ in the cross section [Eq.~(\ref{diffcross})] is less suppressed. 

\item {\it The momentum-averaged cross section.} The calculation of the \n\ optical depth (Sec.~\ref{sec:opt}) requires the convolution of the total cross section $\sigma_Z$ for a given momentum of the background \n\ with the momentum distribution of the \cnb\ [Eq.~(\ref{dnnu})]:
\be
 \bar \sigma_{Z}(E^\prime, T, m_j) =\frac{ \int  \sigma(E', p, m_{j}) dn(p, T)}{\int dn(p, T )}.
\label{momav}
\ee
We have performed this calculation numerically for a wide range of beam energies and \n\ background temperatures.   The results are illustrated in Fig.~\ref{xsectionthermal}.  We see that, as expected, the effect of including the momentum distribution of the background makes the resonant peak smoother and broader: in addition to the broadening due to the angular integration, here the momentum distribution of the background further widens the range of beam energy where the resonance can be realized. 

As the figure shows, the thermal effects on $\bar \sigma_{Z}(E^\prime, T, m_j)$  vary from negligible to substantial as the \n\ mass varies from much larger to comparable to the root mean square of the \cnb\ momentum, $\bar p$, or, for a fixed \n\ mass, as the temperature rises so that $\bar p$  becomes comparable to the mass.  Hence, the thermal effects become important at redshifts larger than $z_{{\rm th},j}$ as in Eq.~(\ref{zd}). 

\end{enumerate}

The momentum-averaged cross section gives  a fully realistic description, that can be compared with some approximate treatments of the problem, shown in Fig.~\ref{xsectionappr}. In order of sophistication, they include: (i) neglecting the \bg\ \n\ momentum altogether, which overestimates the energy of the resonant enhancement (ii) including the \bg\ temperature in the form of an effective \n\ mass $m_{{\rm eff}, j}\simeq \sqrt{\bar p^2 + m^2_j}$, which reproduces the position of the resonance peak, and  (iii) using the total cross section for the \bg\  \n\ momentum fixed at its root mean square value and averaged over the scattering angle. This captures in part the spread of the resonance peak over a range of energies.  This range is further broadened for the full result, $ \bar \sigma$.  
 
In addition to the resonant channel,  the non-resonant ones are considered as well, as summarized in Appendix \ref{app:cs}.  Rigorously, one should repeat the considerations above to calculate the contributions of these channels.  However,  these non-resonant contributions are smooth functions of $s$,  therefore we expect  that the smearing effect due to the \bg\ temperature is well captured by using an averaged value of $s$ instead of the exact expression in Eq.~(\ref{svar}): 
\be\label{bars}
\bar s(E', m_j) \sim 2E'  \sqrt{\bar p^2 + m_j^2}.
\ee
 For simplicity, we use this prescription to calculate the contribution of the non-resonant channels to the total momentum-averaged cross section, $\bar \sigma_{\rm nr}(E^\prime, m_j)=\sum\limits_{i} \sigma_{{\rm nr}, i}(\bar s)$, where $\sigma_{{\rm nr}, i} (s)$ is the non-resonant cross section for a given channel, $i$ (Appendix \ref{app:cs}).

%===================================
\subsection{Optical depth and flux suppression}
\label{sec:opt}

From the results of the previous section, it is immediate to find the scattering rate of a beam \n\ of energy $E'$ and flavor $\alpha$ in a \n\ \bg\  of momentum $p$, whose distribution is given by Eq.~(\ref{dnnu}) (see Appendix \ref{app:amplitude} for derivation): 
\ba
\label{GammaTot}
\Gamma_{\alpha}(E', T )  &=&\sum\limits_{j}  |U_{\alpha j}|^2 \int  \sigma(E', p, m_{j})~ dn(p,T) \nonumber \\
&\equiv& \sum\limits_{j} |U_{\alpha j}|^2~ n(T) ~ \bar \sigma(E', T, m_j),
\ea
where the sum is over all  mass eigenstates, $j$, and $n(T)= \int dn(p, T )$ is the number density of each \n\ species (assumed to be the same for all species). 

The next step is to calculate the optical depth --  i.e., the the total number of collisions of an UHE neutrino of flavor $\alpha$ with the \cnb\ neutrinos throughout its path -- considering that the energy of the beam and the momentum and temperature of the \bg\ undergo redshift.  
For convenience, in what follows $T_0$ is the temperature of the \cnb\ today, and $E$ will represent the energy of the beam \n\ at Earth, i.e., $E = E'/(1+z)$.  In these terms, the optical depth for a \n\ produced at redshift $z$ can be written as: 
\be\label{tau}
\tau_{\alpha}(E, z) = \int_{t(z)}^{t_0} dt' ~\Gamma_{\alpha}[E(1+z'), T_0(1+z')]=  \int_{0}^{z} \frac{dz' }{(1+z') H(z')} ~ \Gamma_{\alpha} [E(1+z'), T_0(1+z')]~.
\ee

Let us estimate for what values of $z$ absorption is significant, i.e., $\tau_\alpha \gta 1$. The total non-resonant cross section at $s \gtrsim m_W^2$ is approximately $\sigma_{\rm nr} \approx \sigma_{tZ} + \sigma_{tW} \approx 7.8 ~G_{F}^2 ~m_{W}^2 / \pi \sim 8.3 \times 10^{-34}~{\rm cm}^2$ [see Eqs.~(\ref{sigmatz}) and (\ref{sigmatw})].   Using this value as a constant in Eq. (\ref{tau}), one gets 
\be
\tau_{\rm nr} \approx 1.0 \left(\frac{1+z}{140}\right)^{3/2}.
\ee
Therefore, provided that $s \gtrsim m_W^2$, neutrinos are completely absorbed at all energies at $z \gtrsim z_{\nu} \approx 140$ due to non-resonant scatterings. Thus, we shall call $z_{\nu}$ as the neutrino horizon beyond which no neutrinos can propagate to us. 
Similarly,  by using the maximum value of the resonant cross section, $\sigma_{\rm r} \sim 5\times 10^{-32}~ {\rm cm}^2$ [Eq.~(\ref{sigmares})], we get an estimate of  the optical depth for the resonant channels:
\be\label{taunr}
\tau_{\rm r} \approx 1.0 \left(\frac{1+z}{10}\right)^{3/2}.
\ee
Thus, resonant absorption occurs if the beam energy is around the resonant energy $E'_{\rm res} \sim m_{Z}^2/ \sqrt{\bar p_0^2 (1+z)^2 + m_j^2}$ at $z \gtrsim z_{\rm dip} \approx 10$.

From the optical depth, we obtain the suppression that applies to a flux of \ns\ $\nu_\alpha$, produced at redshift $z$  and arriving 
at Earth  with energy $E$: 
\be\label{psurv}
P_{\alpha}(E,z) = e^{- \tau_{\alpha}(E,z)}.
\ee
Assuming that UHE \n\ detectors are flavor-blind, the relevant quantity is the suppression of the total flux, defined by the flux-averaged transmission probability:
\be
P(E,z)   \equiv \frac{\sum_{\alpha} \phi_\alpha(E) P_\alpha(E,z) }{\sum_{\alpha} \phi_\alpha(E)}~, 
\label{pave}
\ee
with $\phi_\alpha(E)$ being the flux of \ns\ and antineutrinos of a given flavor $\alpha$ (under the assumption that  \ns\ and antineutrinos have the same  transmission and oscillation probabilities, which is justified for a CP-symmetric \n\ background)\footnote{In fact, the beam neutrinos are in a mixed flavor state, since they oscillate, with oscillation length $L_{\rm beam} = 4\pi E'/ |\Delta m^2|$. These oscillations average out in cosmological distances since $L_{\rm beam}/ H^{-1}  \sim 3.7 \times 10^{-7}~(1+z)^{-5/2}~ [E/(10^{12} {\rm GeV})] ~(|\Delta m_{21}^2|/|\Delta m^2|)  \ll 1$ for all redshifts, where $H$ is the Hubble parameter at epoch $z$. In Appendix \ref{app:amplitude}, it is demonstrated that, as long as the background \n\ is in a mass eigenstate,  the resonant cross section does not depend on the oscillation phase of the beam \n. }. For definiteness, we consider a flavor-democratic composition,  with $\phi_e : \phi_\mu : \phi_\tau = 1 : 1: 1$ at all energies, which is realized in many cases, at least approximately, due to flavor oscillations. This  gives the simplest expression for $P$: 
\be
P(E,z)  = \frac{1}{3}\sum_{\alpha}  P_\alpha(E,z)  ~. 
\label{pave3}
\ee
We note that $P_\mu \sim P_\tau$ are similar due to the structure of the \n\ mixing matrix, with maximal $\numu - \nutau$ mixing. Therefore   a result close to Eq.~(\ref{pave3}) is valid for different values of the ratio  $\phi_\mu/\phi_\tau$,  e.g., for $\phi_e : \phi_\mu : \phi_\tau = 1 : 2: 0$, which expected at production (i.e., before oscillations) for neutrino generation mechanisms via pion decays.  

\begin{figure}[h]
  \centering \includegraphics[width=0.49\textwidth]{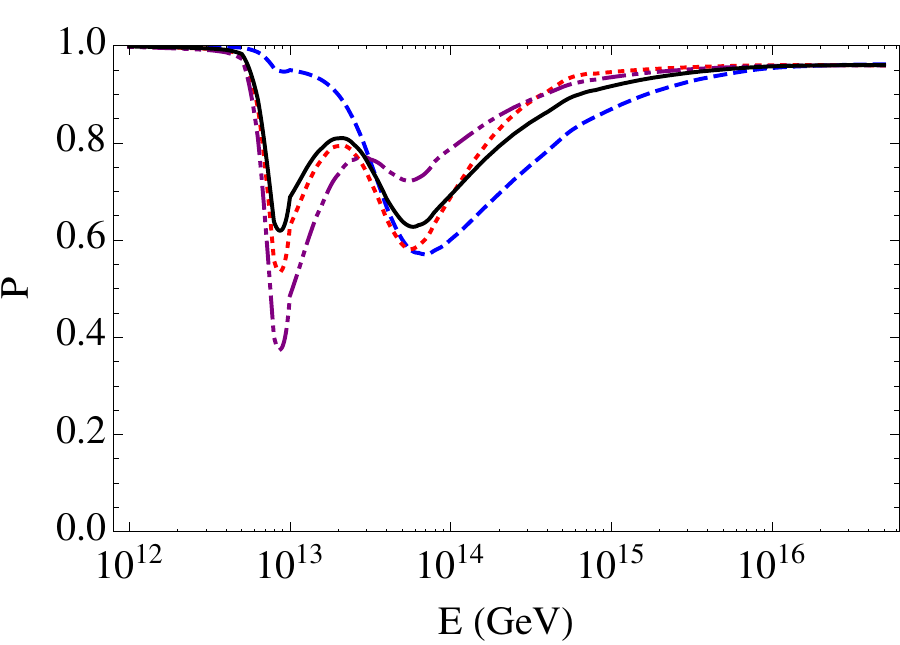}
   \centering \includegraphics[width=0.49\textwidth]{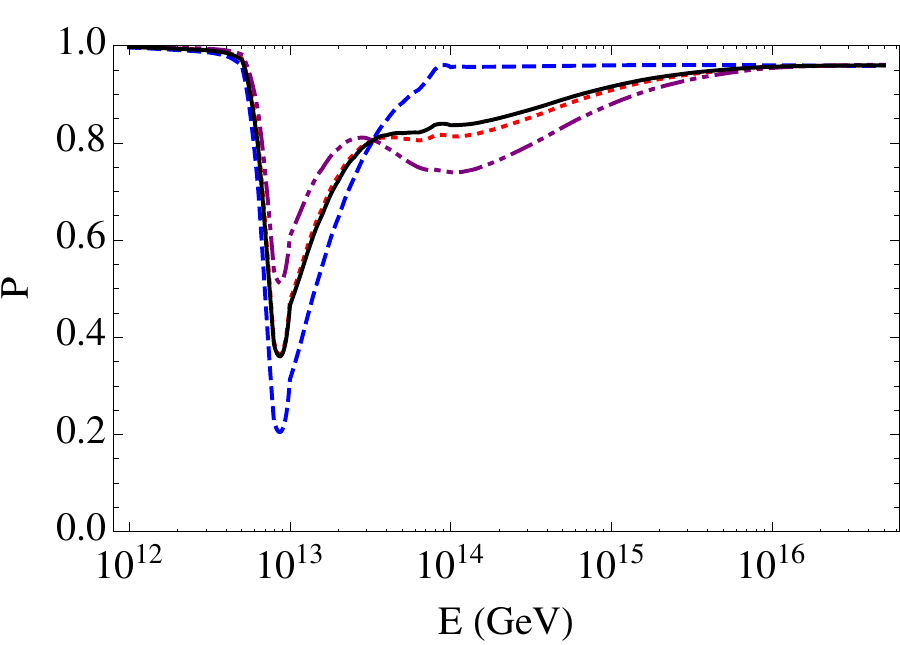}\\
 \caption{Transmission probabilities for UHE neutrinos with different flavors $P_e, P_\mu, P_\tau$ [Eq.~(\ref{pave})] (dashed blue, dotted red, dash-dotted purple),  and the average survival probability, $P$ [Eq.~(\ref{pave3})] (solid black), as a function of the energy, for a source at $z=10$. Left (right) panel is for normal (inverted) hierarchy. The lightest neutrino has mass $m_1~(m_3) = 10^{-5}$ eV.  }
\label{ppartial}
\end{figure}

The main features of $P_e, P_\mu$,  $P_\tau$ and $P$ are shown in Fig.~\ref{ppartial}. In principle, we expect each of these probabilities to exhibit three suppression dips, which we name $D_1, D_2, D_3$, corresponding to the three values of the \n\ masses, $m_1, m_2, m_3$.  The order of the dips with increasing energy is the inverse of the order of masses, therefore  the order is $D_3, D_2, D_1$ for NH and $D_2, D_1, D_3$ for IH.    For $z$ larger than a few, $D_1$ and  $D_2$ appear fused into a single dip ($D_{12}$). This is due to the thermal effects, because the smaller mass gap is comparable with the \n\ average momentum, $m_2 - m_1 \lta 10^{-2}~{\rm eV} \simeq \bar p $ at redshift $z=10$. 
For $P_e$ the dip $D_3$ is suppressed, as a result of  $U_{e3}$ being small (Sec.~\ref{massmix}).    
This explains the dip structure of   the flavor-averaged probability, $P$, and in particular the fact that the lowest energy dip is less deep for NH than for IH. 
%Survival probability for individual flavors and averaged ones we use.

%Survival probabilities
\begin{figure}[t]
  \centering \includegraphics[width=0.42\textwidth]{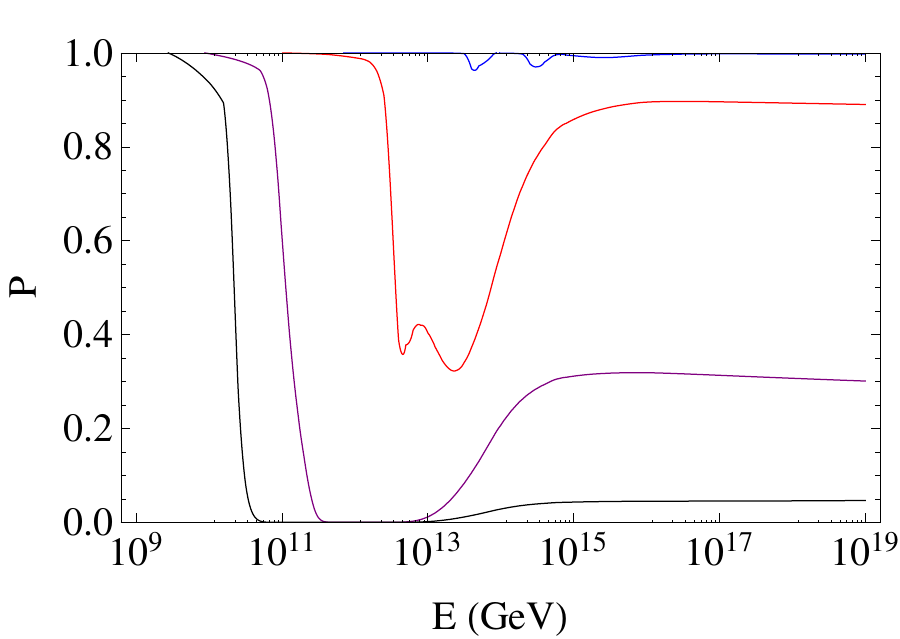}
   \centering \includegraphics[width=0.42\textwidth]{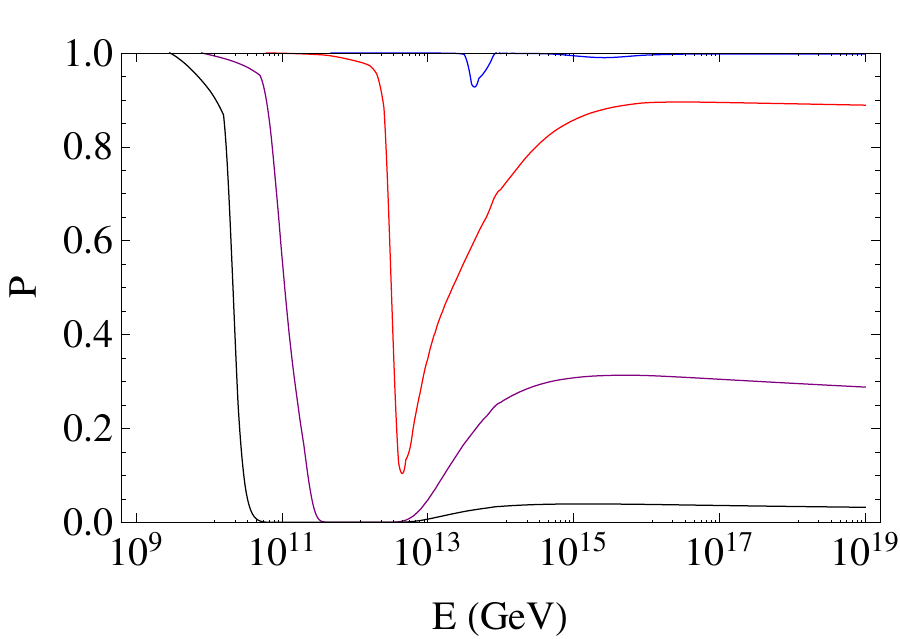}\\
   \centering \includegraphics[width=0.42\textwidth]{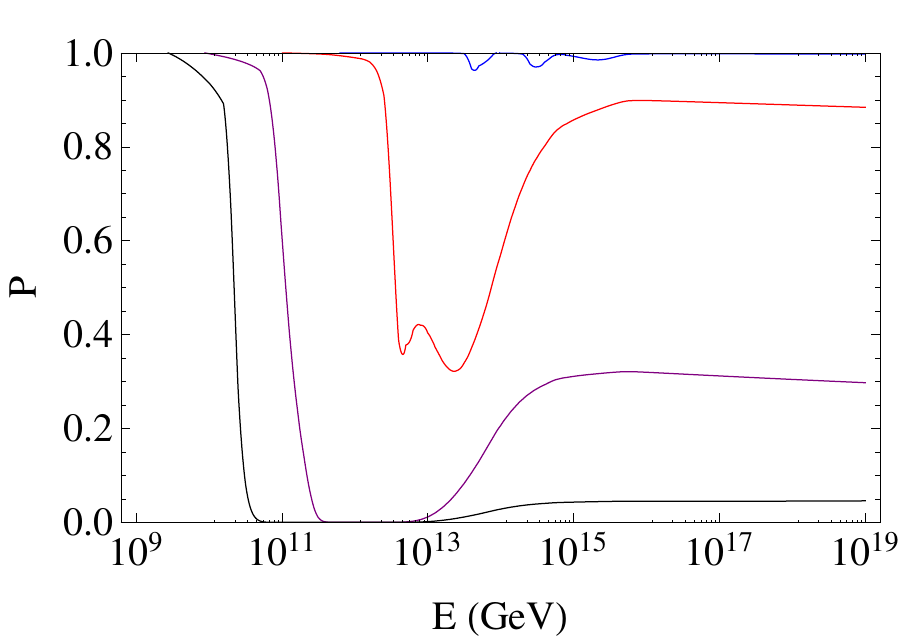}
   \centering \includegraphics[width=0.42\textwidth]{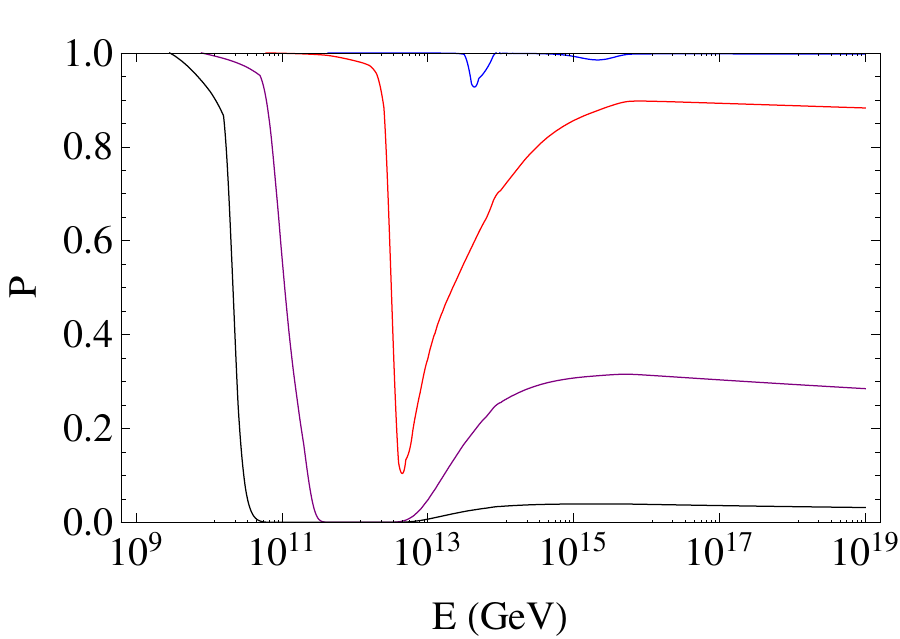}\\
   \centering \includegraphics[width=0.42\textwidth]{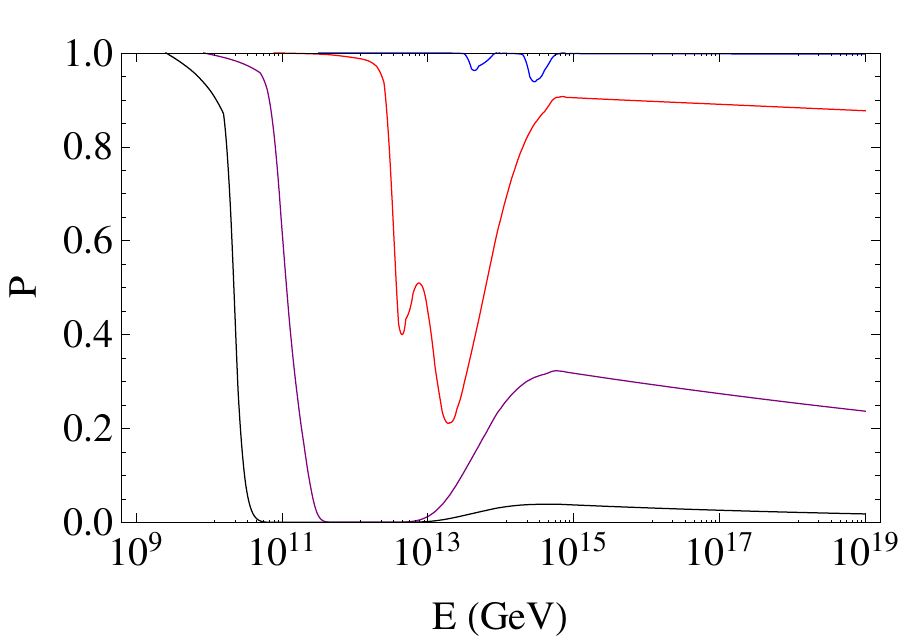}
   \centering \includegraphics[width=0.42\textwidth]{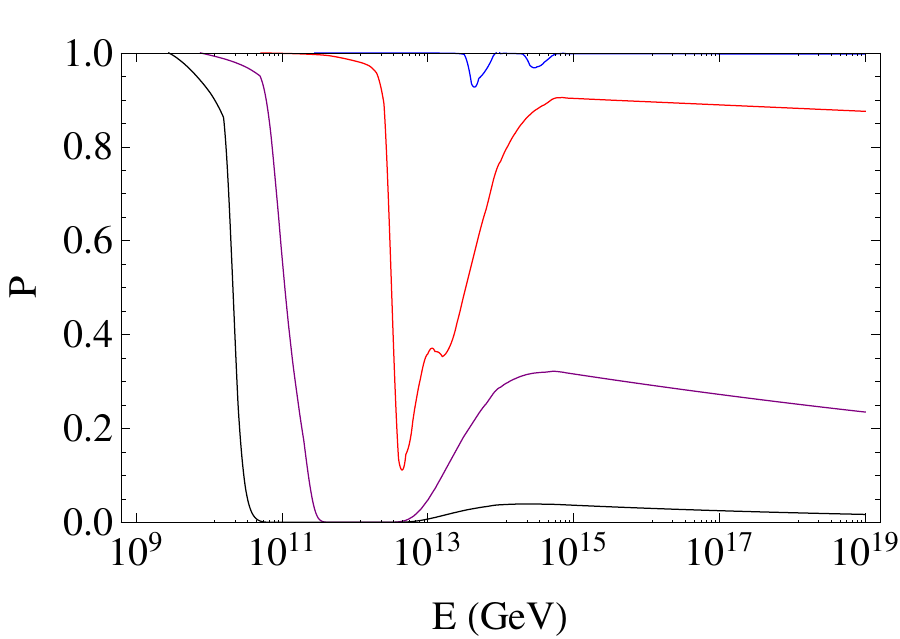}\\
   \centering \includegraphics[width=0.42\textwidth]{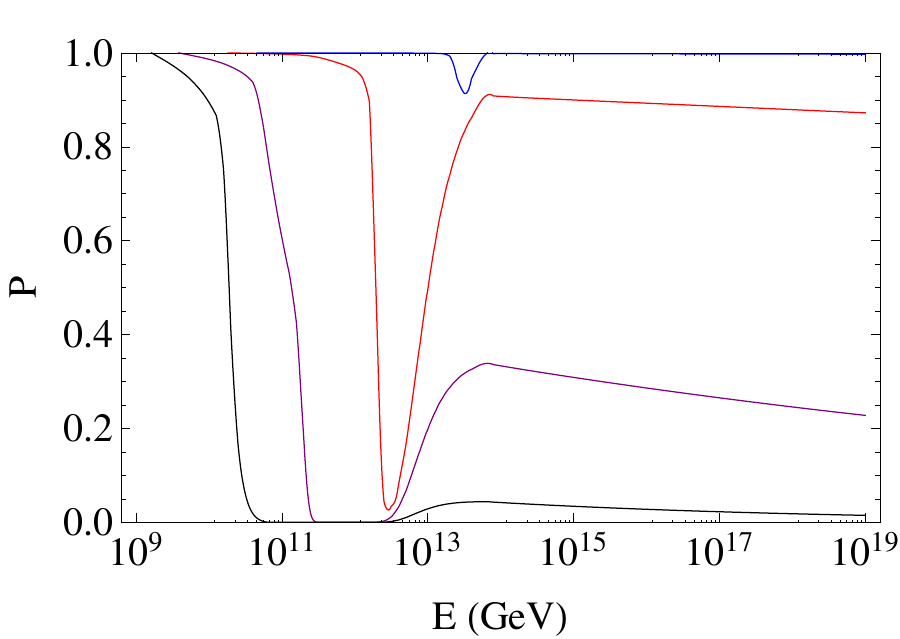}
   \centering \includegraphics[width=0.42\textwidth]{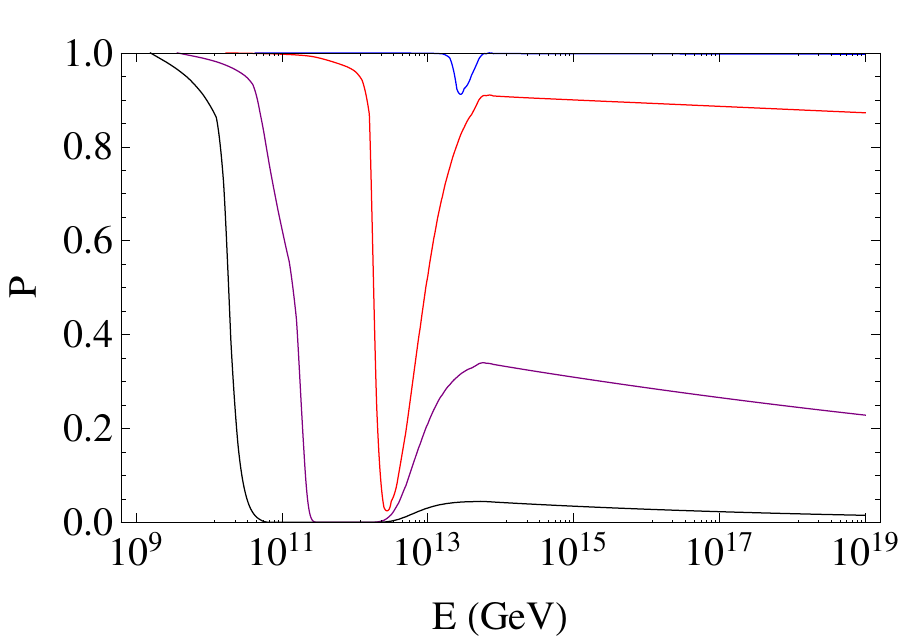}\\
 \caption{Flavor-averaged survival probability given by Eq.~(\ref{pave3}), as a function of the observed neutrino energy, for a source located at $z=1$ (blue), $20$ (red), $100$ (purple), $200$ (black) (curves from top to bottom in each figure). Left (right) column is for normal (inverted) hierarchy. Figures from top to bottom correspond to the lightest neutrino mass $m_1~(m_3)$ in eV: $10^{-5}$, $10^{-3}$, $2 \times 10^{-2}$, $8 \times 10^{-2}$.}
\label{paverage}
\end{figure}

In Fig.~\ref{paverage}, the dependence of $P$ on the energy  and on the production redshift is illustrated in more detail.  The figure shows how resonant absorption becomes substantial ($P \lta 0.5$) for $z\gta 10$, as expected from Eq. (\ref{zd}).  For the same redshift,  absorption above the resonance starts to have some effect, of about 10\%.  For $z \sim 50 -100$, the three absorption dips merge into a single wide suppression well, that spans more than one oder of magnitude in energy; suppression in the regime above resonance ($s \gta m^2_W$) is of more than 50\%.  Finally, for $z \sim 200$, suppression is nearly complete at $E \gta 10^{11}$ GeV, where the non-resonant contribution to the cross section alone is enough to have $\tau(E,z)> 1$.

%========================================================================================
\section{Ultra high energy neutrino flux}
\label{sec:uhen}

Excluding the fortuitous case of a \n\ source near Earth,  UHE \ns\ are expected to be detected in the form of a diffuse flux from all the sources in the universe.  To calculate this flux, it is necessary to model the number of sources per comoving volume, per unit of physical time, $t$:
\be
\eta(z) \equiv \frac{1}{r^2}\frac{d^3 N_{\rm s}}{d\Omega dr dt}~,
\ee
and  the \n\ flux  from a single source: 
\be
\phi(E') \equiv \frac{d N_{\nu}}{dE'}~.  
\ee
The product of the two gives the emissivity of an ensemble of sources:
\be
 \mathcal{L}_{\nu}(E^\prime, z)=\eta(z) \phi(E')  ~,
\ee
from which one gets  the diffuse flux (i.e., the number of neutrinos per unit energy per unit area per unit time per solid angle), in terms of the energy of the \ns\ at Earth  $E = E'/(1+z)$ \cite{Eberle:2004ua}:
 \be
J_{\nu}(E) =   \frac{1}{4\pi} \int_{0}^{\infty} \frac{dz}{H(z)} ~ P(E,z)~ \mathcal{L}_{\nu}[E(1+z), z],
 \ee
where Eq.~(\ref{dr}) was used, and $P(E,z)$ is the average transmission probability given by Eq. (\ref{pave3}). 

Due to the integration over redshift, the suppression of the diffuse flux is less rich of structures  compared to the case of a single source at fixed redshift.  Therefore, we expect that only a single, wide suppression dip will appear in the \n\ spectrum.  This suppression should be stronger for sources whose distribution extends to high redhshifts, $z \gta 10$, where $P \lta 0.5$ in the resonance region (see Fig.~\ref{paverage}).   In the next section, a number of specific examples are discussed. 

%===================================================
\subsection{Top down neutrino sources}

Top down sources are objects predicted from physics beyond the standard model, that radiate light particles in a variety of mechanisms. 
The examples discussed here are  topological defects, such as cosmic strings and cosmic necklaces, and unstable superheavy particles.

%=======================
\subsubsection{Cosmic strings}

Cosmic strings are predicted in field theory models with spontaneous symmetry breaking whose vacuum manifold is not simply connected, e.g., Nielsen-Olesen topological strings in the Abelian Higgs model \cite{Nielsen:1973cs} (for reviews see, e.g., Refs.\cite{VilenkinBook,Copeland:2009ga,Polchinski:2004ia,Copeland:2011dx}). Cosmic F- and D-strings of superstrings theory may also be produced in the brane inflation models in string theory \cite{Sarangi:2002yt}. If they exist, cosmic strings are  stable relics formed in the very early universe, thus, they have incredibly high energy densities in their core. They are characterized by their tension, $\mu$ (mass per unit length) denoted in Planck units as a dimensionless parameter $G\mu$, where $G$ is Newton's constant. The upper bound on cosmic string tension from CMB anisotropy measurements of WMAP and SPT is $G\mu \lesssim 1.7 \times 10^{-7}$ \cite{Dvorkin:2011aj}, and it has recently been updated by Planck to $G\mu \lesssim 1.5\times 10^{-7}$ \cite{Ade:2013xla} which corresponds to a mass scale  $m_s \lta \sqrt{\mu} \sim 5 \times 10^{15}$ GeV. This suggests that cosmic strings may be responsible for extremely high energy cosmic rays in the universe if they can emit particles efficiently. 

Various mechanisms to produce particles from cosmic strings have been studied \cite{Hill:1986mn,Bhattacharjee:1989vu,MacGibbon:1989kk,Brandenberger:2009ia,Vachaspati:2009kq,Berezinsky:2009xf,Berezinsky:2011cp,Lunardini:2012ct}, but only a few of them yield observable fluxes. For instance, observable UHE neutrinos can be achieved at the cusps of superconducting cosmic strings \cite{Berezinsky:2009xf}, and at the cusps \cite{Berezinsky:2011cp} and kinks \cite{Lunardini:2012ct} of cosmic strings and cosmic superstrings. Recently, Kaluza-Klein mode emission from cosmic superstring cusps has been shown to be an efficient radiation mechanism \cite{Dufaux:2011da,Dufaux:2012np}, which can also lead to UHE neutrinos, and can be a potentially interesting signature of superstring theory.

In what follows, as an example, we discuss the case of \n\ emission from cosmic string cusps and kinks via heavy scalars (moduli) \cite{Berezinsky:2009xf,Berezinsky:2011cp,Lunardini:2012ct}. 

The neutrino emissivities from cusps \cite{Berezinsky:2011cp} and kinks \cite{Lunardini:2012ct} (via modulus emission from cosmic strings) are respectively given by:
\ba
\mathcal{L}_{\nu}^{\rm cusp} &=& 9.5\times 10^{23} \frac{\alpha^2 (G\mu)^{1/2} \ln[(G\mu)^{1/2} m_p/m] }{p (1+z)^{5}} \frac{m_p}{E^2 t_p^{1/2} t(z)^{7/2}}, \label{cusp}
\\
\mathcal{L}_{\nu}^{\rm kink} &=& 1\times 10^{23} \frac{\alpha^2 (m_p/m)^{1/2}}{p (1+z)^{5}} \frac{m_p}{E^2 t(z)^{4}}, \label{kink}
\ea
where $m_p$ is the Planck mass, $t_p$ is the Planck time, $t(z)$ is the cosmic time given by the integral of Eq.~(\ref{dt}) from epochs $z$ to $\infty$, $p \lesssim 1$ is the string reconnection probability, $G\mu$ is the string tension, $m$ is the modulus mass and $\alpha$ is the modulus coupling constant.  In both models, the \n\ production has a redshift cutoff, 
\be
z_{\rm min}^{\rm str}  \sim 122 \left(\frac{G\mu}{10^{-17}}\right)^{2/7}  \left(\frac{m}{10^4~ \rm GeV }\right)^{2/7}  \left(\frac{E}{10^{11}~ \rm GeV}\right)^{-4/7}~,
\ee
that corresponds to the minimum energy at which the hadronic cascade produces pions ($\epsilon \sim 1$ GeV in the rest frame of the modulus), therefore the expressions above are valid for  $z> z_{\rm min}^{\rm str}$.   Eqs.~(\ref{cusp}) and (\ref{kink}) show that in both cases  the emissivity is dominated by the emission at low redshifts, therefore the suppression of the diffuse flux due to \abs\ should be  roughly  determined by $P(E,z^{\rm str}_{\rm min})$. 
Here, we used the following parameter values: for kinks, $\alpha \sim 1$, $m \sim 10^4$ GeV, $G\mu \sim 10^{-17}$ and $p \sim 1$, corresponding to $z_{\rm min}^{\rm str} \simeq 2.3-122$ in the interval $E \simeq 10^{11} - 10^{14}$ GeV; for cusps, $\alpha \sim 2\times 10^7$, $m \sim 10^4$ GeV, $G\mu \sim 6\times 10^{-19}$ and $p \sim 1$, which give $z_{\rm min}^{\rm str} \simeq 1.0- 54$ for  $E \simeq 10^{11} - 10^{14}$ GeV. 

Fig.~\ref{flux_top} shows the diffuse flux from cosmic string kinks and cusps, with \abs\ effects, for the eight \n\ mass spectra (four for each hierarchy) listed in Table \ref{tabmass}.   The flux has a sharp cutoff at about $E \sim 10^{10}-10^{11}$ GeV. This is where the  $z_{\rm min}^{\rm str} \sim z_{\nu}\sim 140$, so that the entire flux is emitted beyond the \n\ horizon $z_\nu$, and is completely absorbed before reaching Earth. 
In the spectrum, we observe the expected smearing of the dips into a single, broad suppression feature in the energy interval $E \sim 10^{11} - 10^{14}$ GeV.  The suppression is overall stronger for kinks, due to the higher values of $z_{\rm min}^{\rm str}$.   The dependence of the suppression on the \n\ mass spectrum is fairly weak: the spectrum shape is nearly identical for the 
all cases except for the one with the largest mass.  This is due to a combination of the  two smearing effects discussed above, due to redshift integration and to the thermal effects.  Considering large values of $z^{\rm str}_{\rm min}$, thermal effects  influence the position and depth of the dips more than the \n\ mass itself, at least for the strongly hierarchical \n\ spectra.  

For superconducting string cusps, the neutrino emissivity  is given by \cite{Berezinsky:2009xf}
\be
\mathcal{L}_{\nu}^{\rm sup} = 1.4 \times 10^{22}  \frac{i_c f_{B}}{(1+z)^{5/2}} \frac{B m_p t_p^{1/2}}{E^2 t(z)^{5/2}},
\ee
where $i_c \lesssim 1$ is the dimensionless string parameter characterizing the maximum current on the string, $f_{B} \sim 10^{-3}$ is the magnetic field filling factor, $B \sim 10^{-6}$ G is the magnetic field strength. In Fig.~\ref{flux_nodip}, we take $i_c \sim 0.1$.  Like in the previous case, the emissivity is dominated by low redshifts, and has a lower redshift cutoff,
\be
z_{\rm min}^{\rm sup} \sim 1.2 ~i_{c}^{3/2} \left(\frac{G\mu}{6.7\times 10^{-19}}\right)^{-3/4} \left(\frac{B}{10^{-6}~ \rm G}\right)^{2} \left(\frac{E}{10^{-12} ~\rm GeV}\right)^{-3/2}~,
\ee  
furthermore,  $ z <z_{\rm max} \sim 5$, because in Ref.~\cite{Berezinsky:2009xf}, it was assumed that the magnetic fields trace galaxies and clusters, and thus strings have no current at times prior to structure formation. 

In Fig.~\ref{flux_nodip}, we plot the neutrino flux from the superconducting cosmic string cusps. It can be clearly seen that the absorption dips are too tiny to be observable.  Because the dominant redshift $z_{\rm min}^{\rm sup}$ is small,  the optical depth is much less than unity, hence absorption in only at the level of 10\% or less.   Similarly to  cosmic string cusp and kinks, the flux vanishes at about $E \sim 10^{11}$ GeV, when $z_{\rm min} \sim z_{\nu}$.

%==========================
\subsubsection{Cosmic necklaces}

Cosmic necklaces are topological defects made up of strings and monopoles \cite{Berezinsky:1997kd,Berezinsky:1997td}. They are predicted in field theory models, where the symmetry breaking sequence has the form $G \to H \times U(1) \to H \times Z_{2}$, where $G$ is a semi-simple Lie group. As a result of the first symmetry breaking, monopoles form, and after the $U(1) \to Z_2$ breaking, each monopole is attached to two strings, each of which carries out half unit of flux as a result of the remaining $Z_{2}$ symmetry, hence the name cosmic necklace. As the monopoles and antimonopoles on loops of necklaces meet, they annihilate and produce heavy $X$-bosons related to the corresponding symmetry breaking scales of monopoles or strings. The bosons then decay via hadronic cascades into pions, that eventually decay producing numerous UHE neutrinos. 

The neutrino emissivity from cosmic necklaces is given by \cite{Berezinsky:1997td} (see however Ref.~\cite{BlancoPillado:2007zr})
\be\label{necklace}
\mathcal{L}_{\nu}^{\rm neck} = \frac{\Theta[m_X - E(1+z)] e^{- E(1+z)/m_X} }{2 \ln[m_X/(1 \rm GeV)] (1+z)^6} \frac{r}{ E^2 m_p t_p t(z)^3},
\ee 
where $m_X$ is the mass of the emitted heavy boson, and $r$ is a parameter that depends on the monopole mass and the string tension.  The model has a minimum redshift of \n\ emission, $z_{\rm min}^{\rm neck}$, which depends on the lifetime of the necklace. There is also a maximum energy cutoff, where $E' = E(1+z) \sim m_X$; the flux vanishes beyond  this point.  Here we take $m_X \sim 5\times 10^{15}$ GeV, $r \sim 2\times 10^{30}$ GeV$^2$ and $z_{\rm min}^{\rm neck} = 10$.

Fig.~\ref{flux_top} shows the  diffuse flux expected in this model, with \abs\  included for the eight mass configurations of Table \ref{tabmass}. The flux suppression effect is similar to the case of cosmic string cusps and kinks: all these models share the common feature of a large redshift cutoff, $z_{\rm min} \gta 10$, which controls the degree of \abs.  We note, however, that  the cutoff is parameter-dependent: smaller values of $z_{\rm min}$ (i.e., longer lifetime of the necklace) are allowed, and would result in weaker suppression.   
Even for large $z_{\rm min}^{\rm neck}$, the flux from cosmic necklaces may show no absorption effects, if
$m_X \lta 10^{12}$ GeV, which means that the maximum \n\  energy cutoff is  below the range of energy where \abs\ is relevant.

%==============================
\subsubsection{Superheavy dark matter}

The existence of  dark matter is substantiated by various cosmological and astrophysical observations. Although we know the contribution of the dark matter particles to  the cosmic energy density, $\Omega_{\rm CDM} = 0.227$ \cite{Komatsu:2010fb} (expressed as a fraction of the critical density), their fundamental properties (mass, lifetime, spin, etc.) are still largely unknown, and are being tested via indirect and direct detection methods. Considering  that the standard model has a zoo of particles of 3  families, it is natural to imagine that the dark matter sector may consist of multiple particle species.  In this context, there could exist a long lived superheavy dark component -- let's call it $X$ -- with small abundance, i.e., $\Omega_{\rm SHDM} \equiv \xi_{X} \Omega_{\rm CDM}$ with $\xi_{X} \ll1$ \cite{Berezinsky:1991aa,Berezinsky:1997hy,Kuzmin:1997cm,Chung:1998rq,Esmaili:2012us}. The $X$ particles may have masses up to of order GUT scale, $m_{X} \sim 10^{16}$ GeV.  For them to be the thermal relics in the universe, $m_X$ should be less than the reheating temperature, which can be constrained by the tensor to scalar ratio upper bounds to be around $T_{\rm rh} \lesssim 5 \times 10^{15}$ GeV \cite{Ade:2013lta}.  The life time  $\tau_X$ of these objects can be as long as the age of the universe.  However, if they decay via hadronic cascades somewhere between the \n\ horizon and today, a flux of UHE neutrinos should be generated. 

The neutrino emissivity from decaying superheavy dark matter (SHDM) particles is given by \cite{Berezinsky:1997hy}
\be\label{shdm}
\mathcal{L}_{\nu}^{\rm SHDM} = \frac{3 r_x \Omega_{\rm CDM}}{16 \pi} \frac{\Theta[m_X - E(1+z)] e^{- E(1+z)/m_X}} {\ln[m_X/(1 \rm GeV)] (1+z)^2} \frac{m_p H_{0}^2}{E^2 t_{0} t_{p}},
\ee
where $r_x \equiv \xi_{X} t_0/ \tau_X$. Here we use $m_X \sim 5 \times 10^{15}$ GeV, $r_x \sim 3 \times 10^{-7}$. The model has a minimum redshift, corresponding to the time where most of the $X$ particles have decayed (assuming that their lifetime is shorter than the age of the universe). We use $z_{\rm min} =10$.

\begin{figure}[h]
\centering \includegraphics[width=0.4\textwidth]{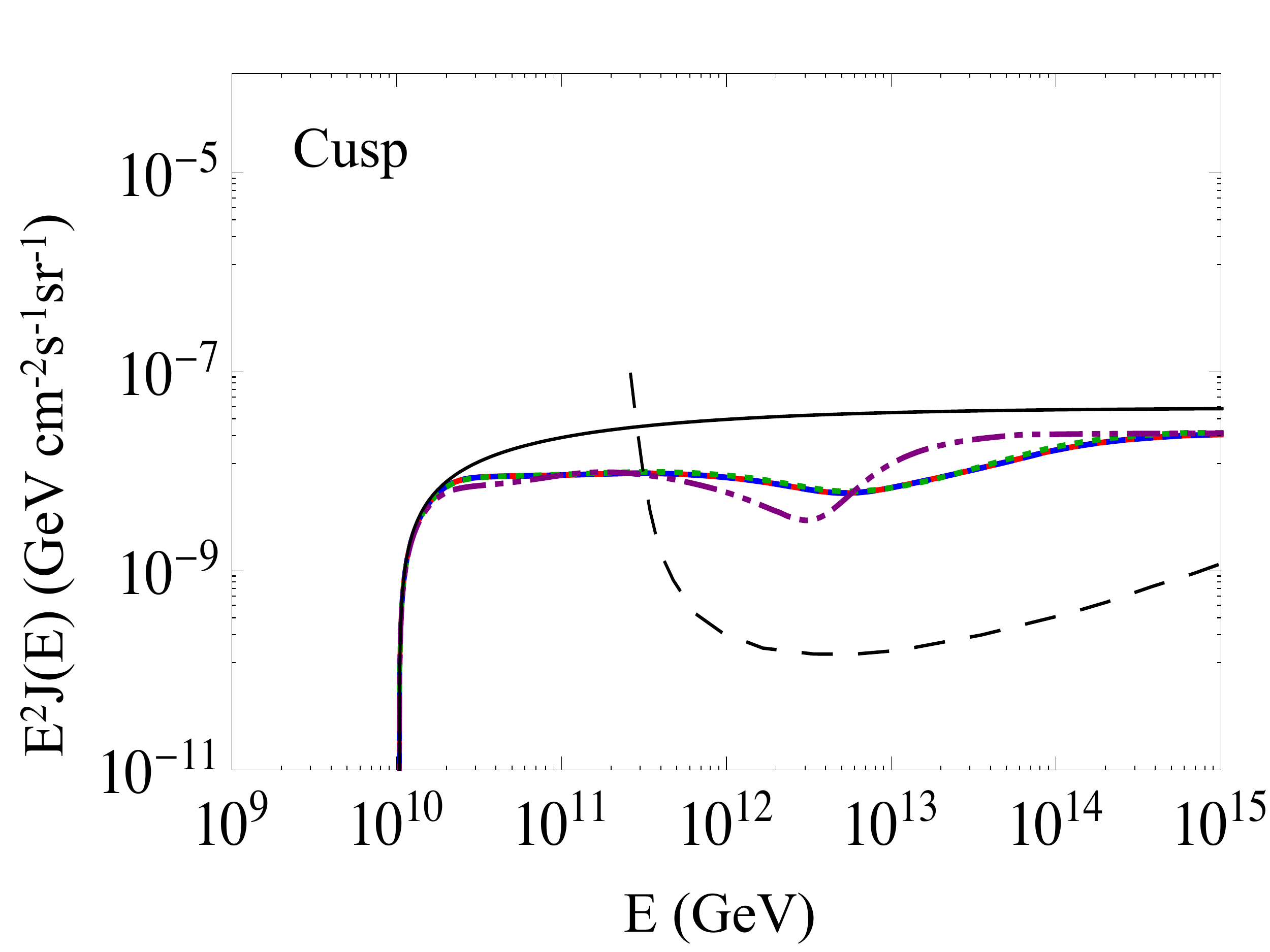}
\centering \includegraphics[width=0.4\textwidth]{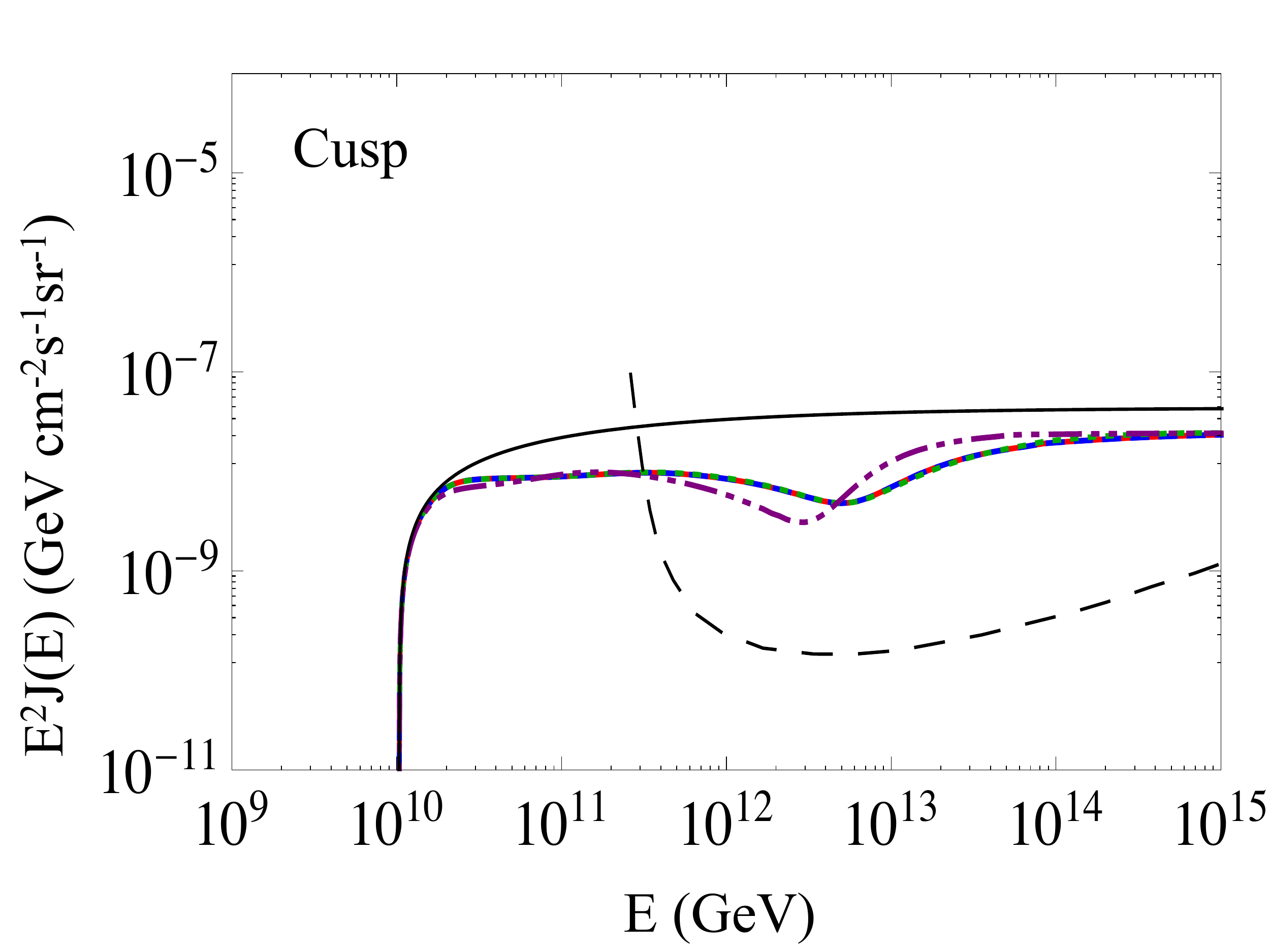}\\

\centering \includegraphics[width=0.4\textwidth]{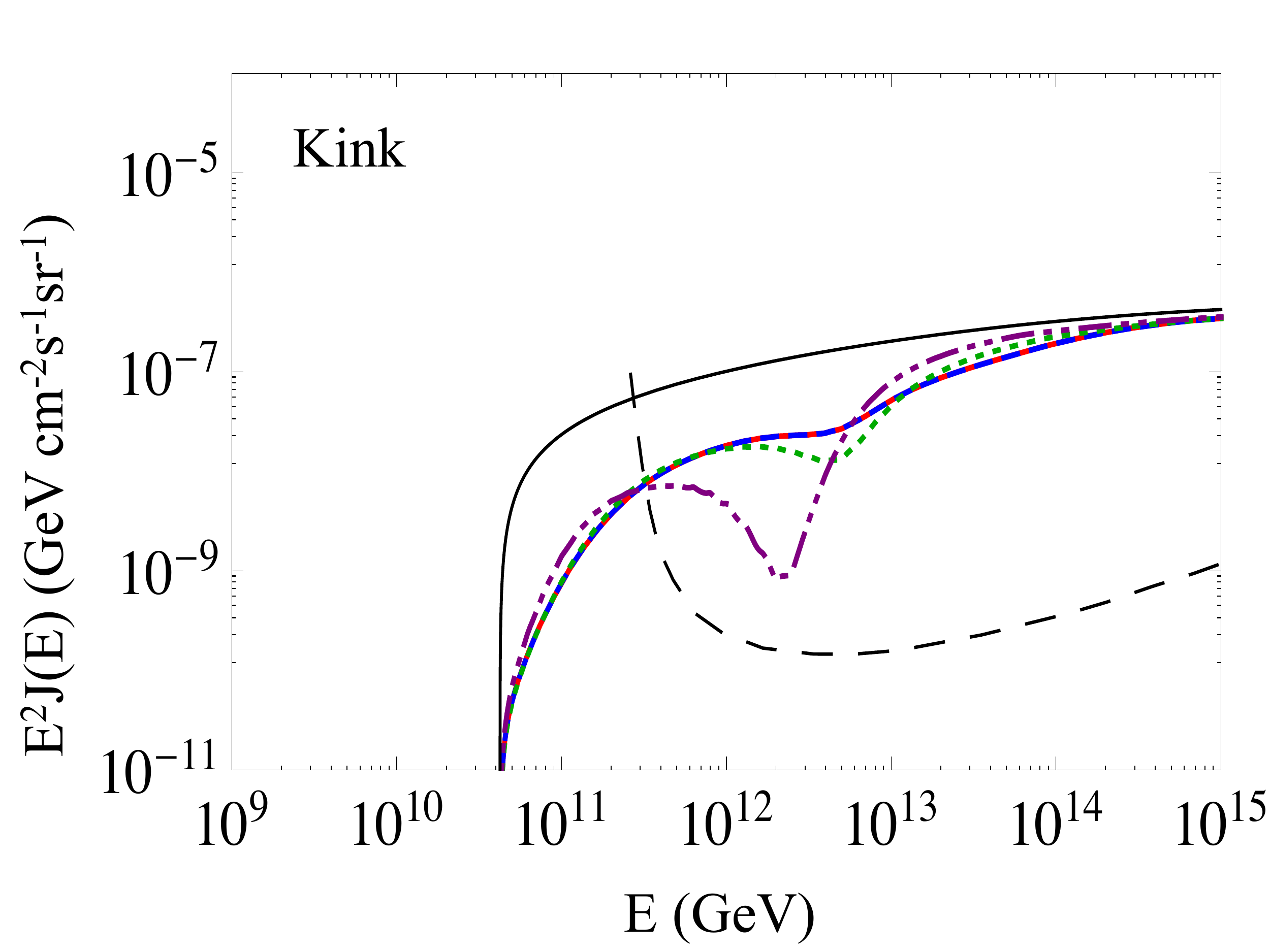}
\centering \includegraphics[width=0.4\textwidth]{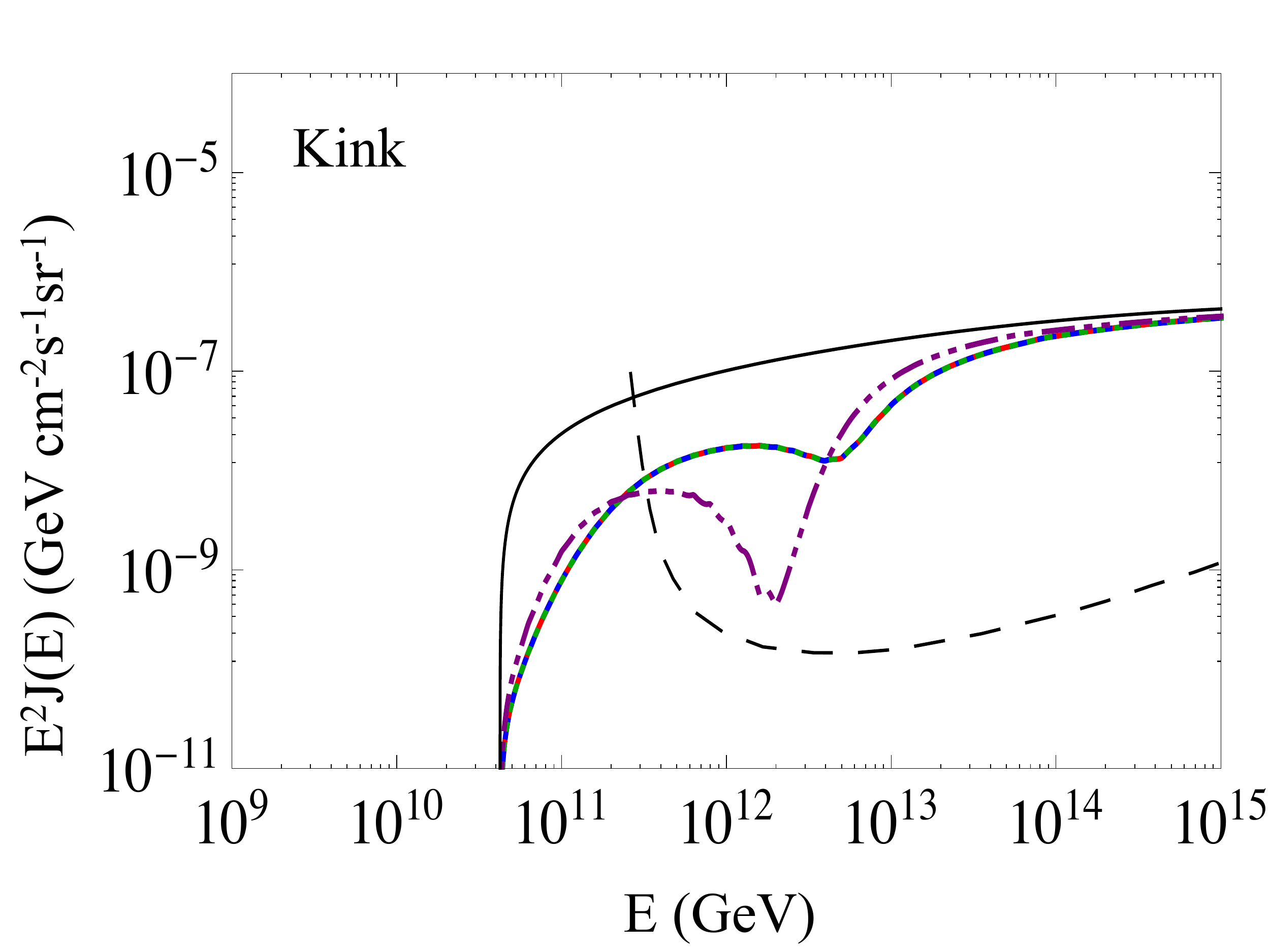}\\
  
\centering \includegraphics[width=0.4\textwidth]{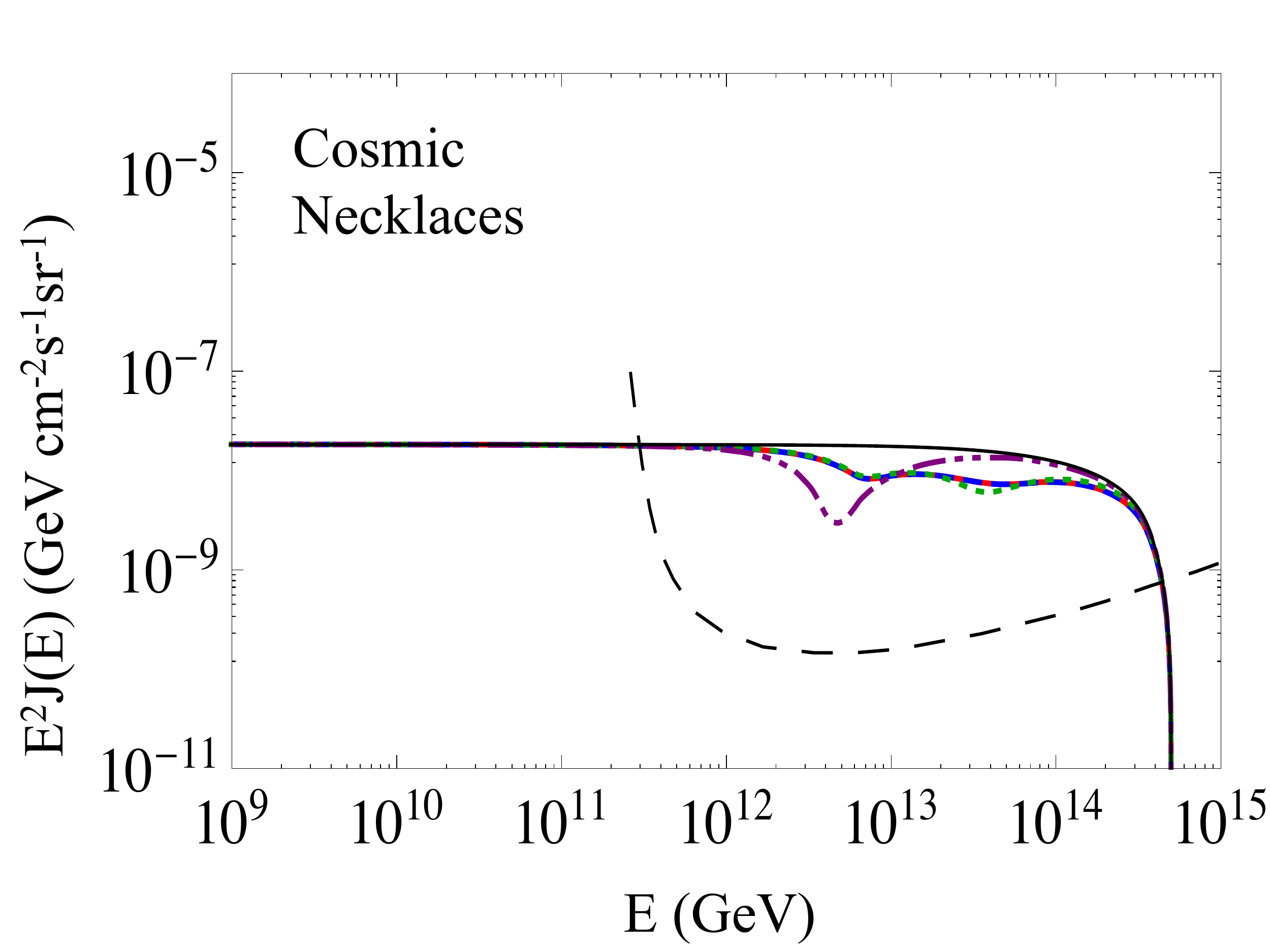}
\centering \includegraphics[width=0.4\textwidth]{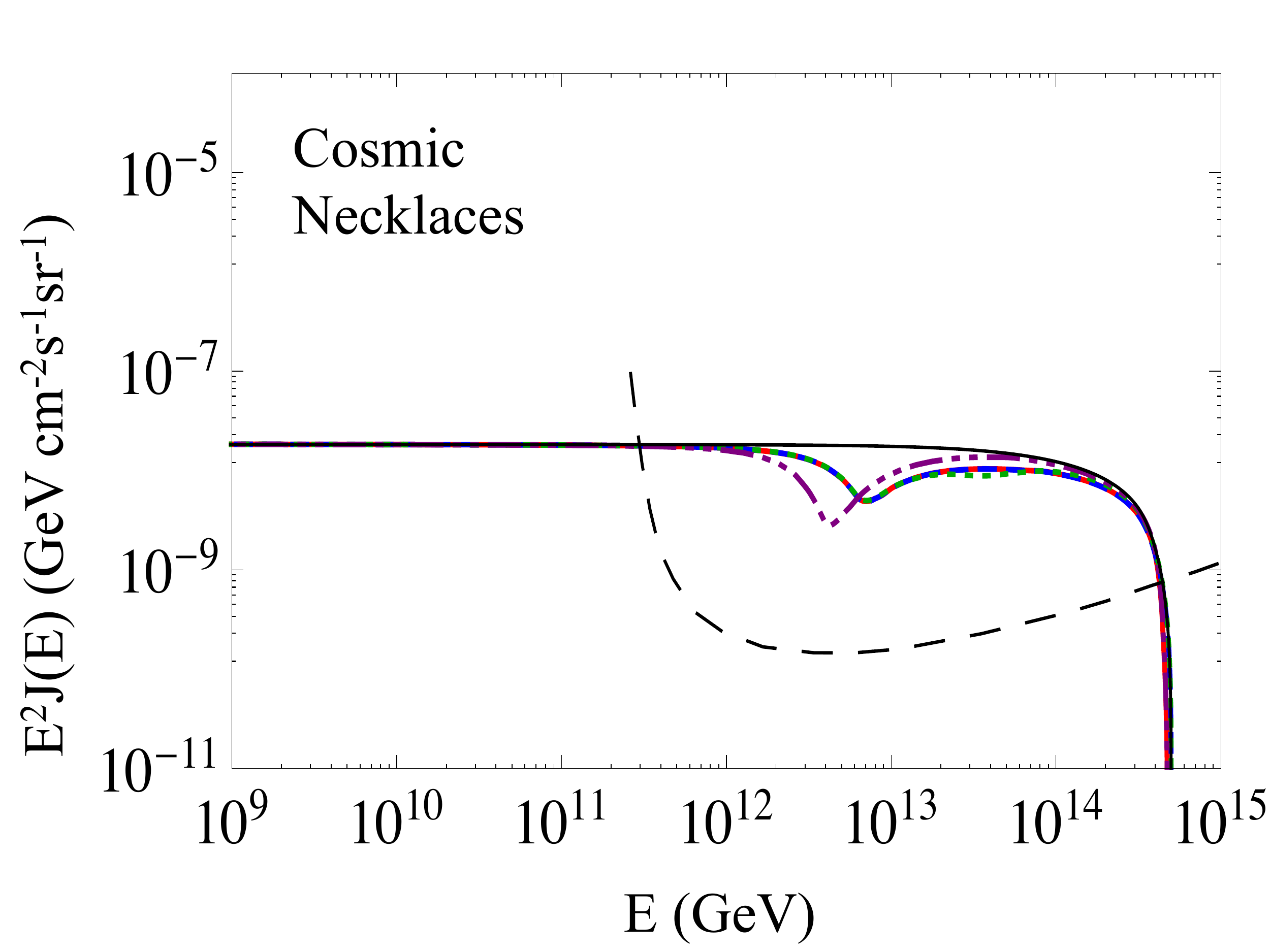}\\

\centering \includegraphics[width=0.4\textwidth]{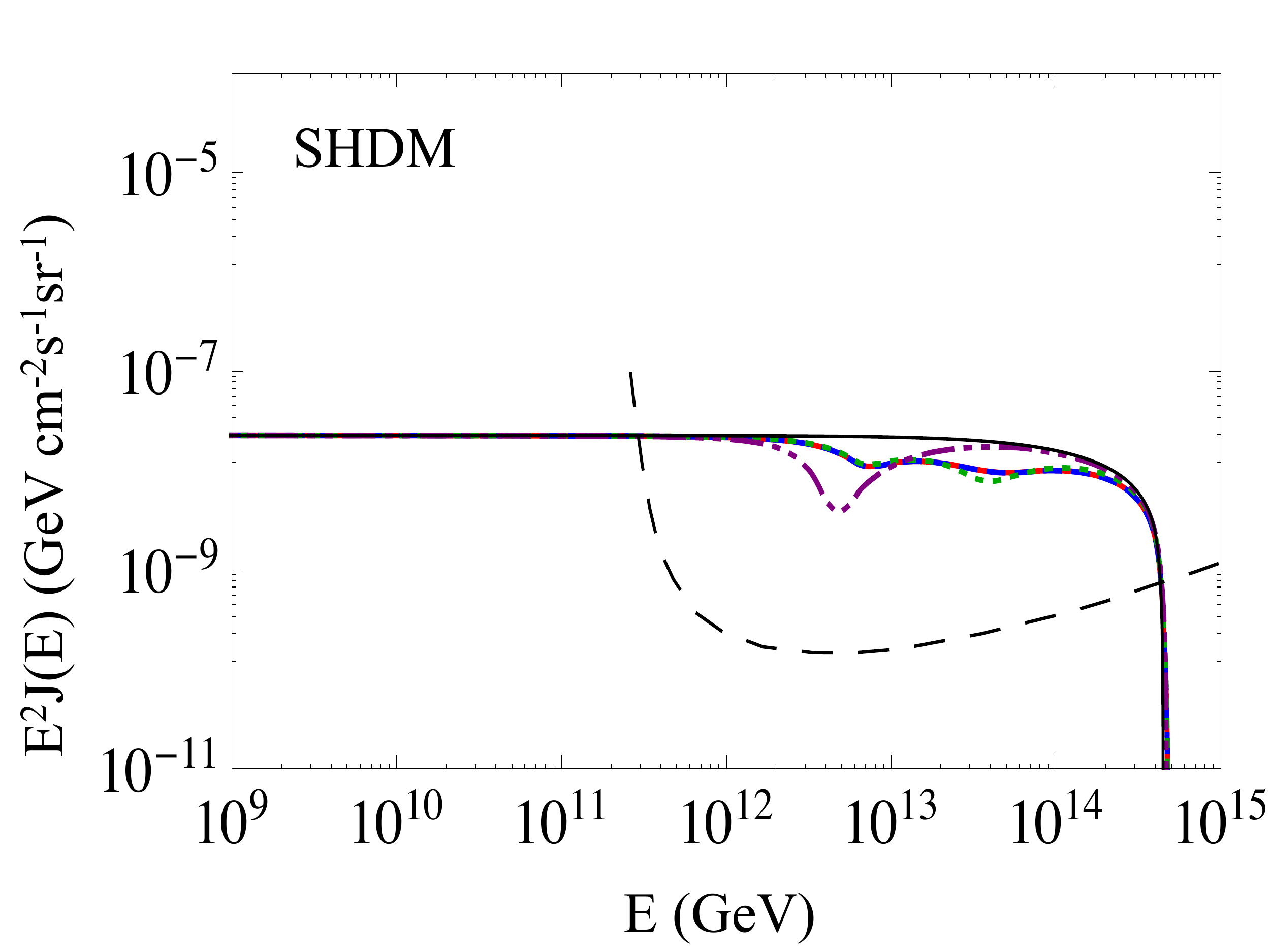}
\centering \includegraphics[width=0.4\textwidth]{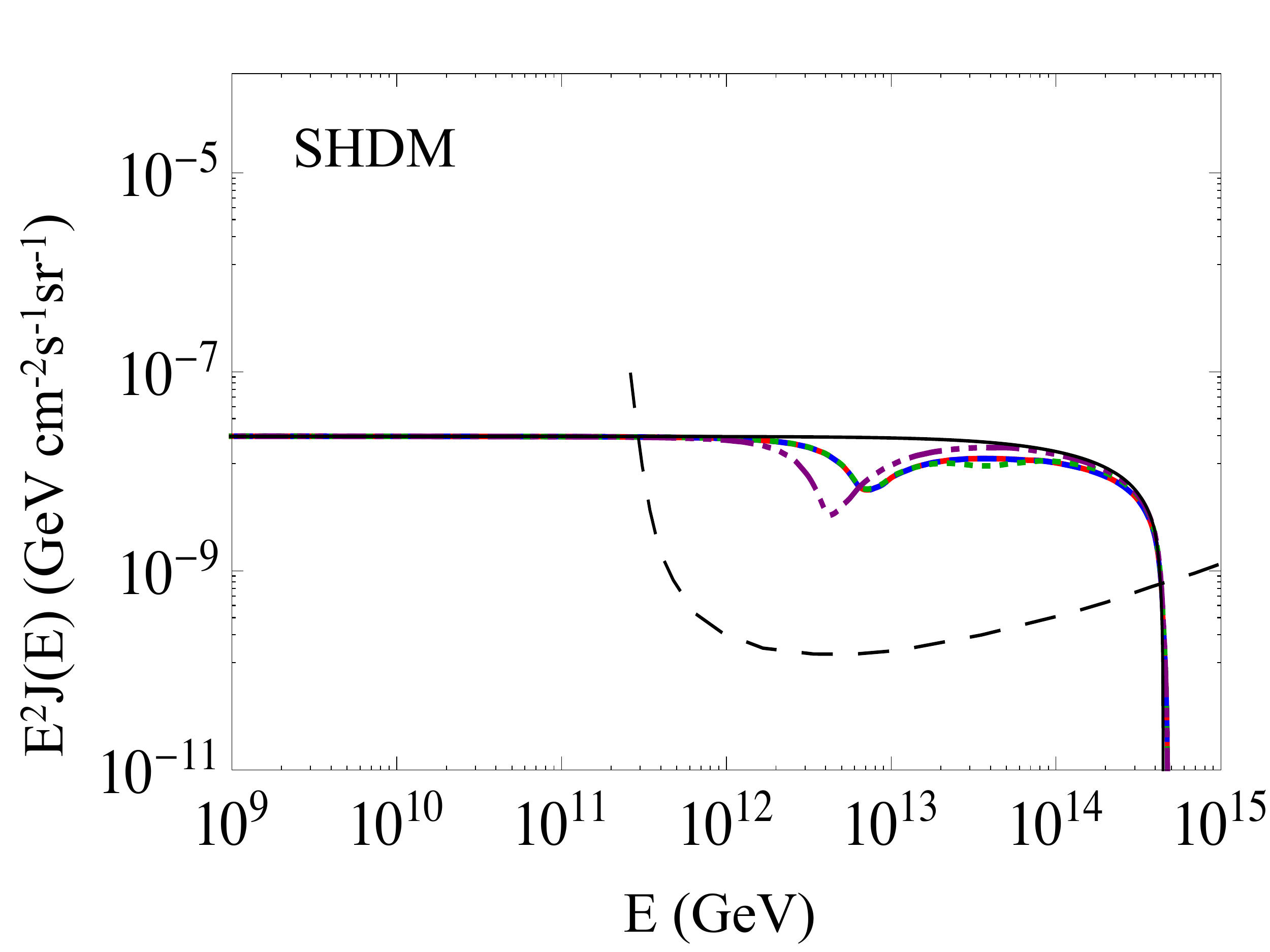}\\
\caption{Expected \n\ fluxes from top down models: cosmic string cusps, kinks, cosmic necklaces and superheavy dark matter (SHDM) (labels in the figures), as a function of energy, for the neutrino masses given in Table \ref{tabmass}. The curves correspond to different values of the minimum neutrino mass $m_{\rm min}$ in eV: $10^{-5}$ (red, solid), $10^{-3}$ (blue, dashed), $2\times 10^{-2}$ (green, dotted), $8\times10^{-2}$ (purple, dot-dashed). The left (right) column is for normal (inverted) hierarchy. Also shown are the fluxes with no absorption (thin, black solid) and the expected sensitivity of SKA (thin, black dashed). } 
\label{flux_top}
\end{figure}

\begin{figure}[h]
\centering \includegraphics[width=0.4\textwidth]{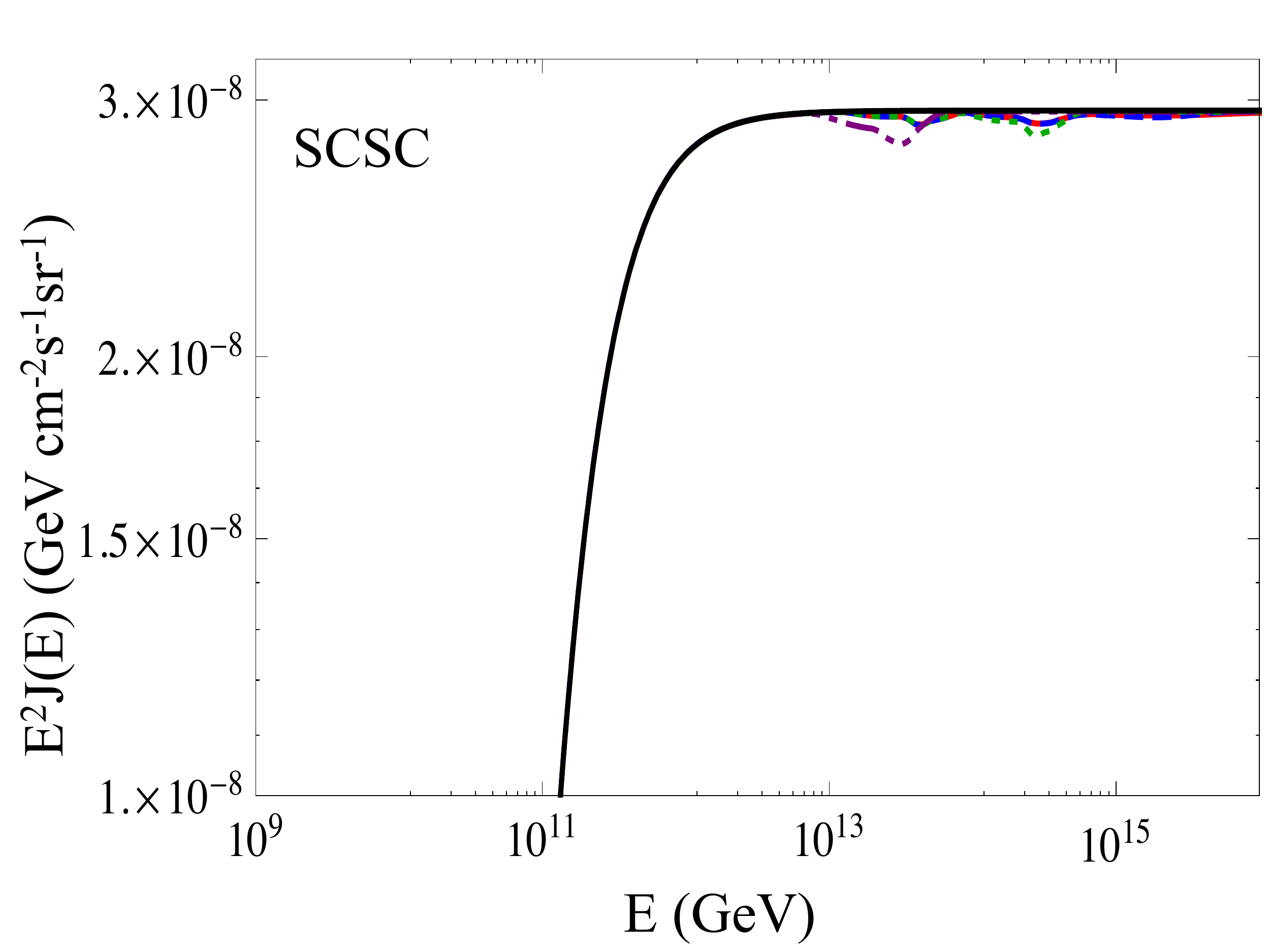}
\centering \includegraphics[width=0.4\textwidth]{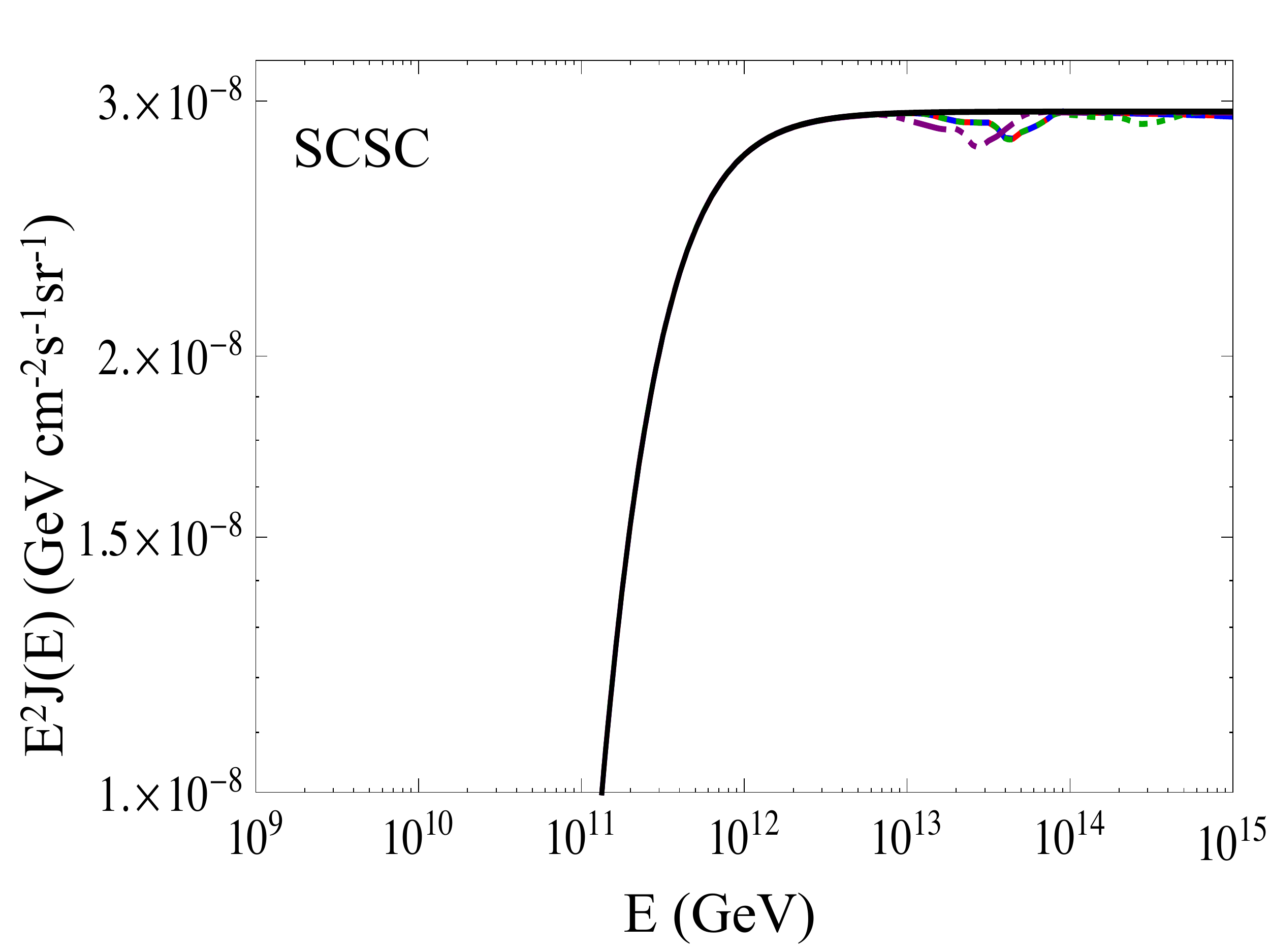}\\
 \caption{Expected \n\ flux from superconducting cosmic string cusps (SCSC) as a function of energy, for the same neutrino mass values (and color coding) as in Fig.~\ref{flux_top}. Left (right) column is for normal (inverted) hierarchy.}
\label{flux_nodip}
\end{figure}

Fig.~\ref{flux_top} shows the resonant dips for different neutrino masses for SHDM whose emissivity is given by Eq.~(\ref{shdm}). Except for the different redshift dependence, this model is very similar to the cosmic necklace model since in both cases a heavy particle decays at rest,  producing neutrinos via hadronic cascades.  The same considerations done for cosmic necklaces apply here, and will not be repeated. 

%Top down models with dips

%Examples with no dips

%=================================
\subsection{Astrophysical neutrino sources}

%=============================
\subsubsection{Cosmogenic neutrinos}

Cosmogenic \ns\ are a guaranteed result of UHE protons scattering off the background of (mostly CMB) photons \cite{Beresinsky:1969qj}.  This is the same phenomenon as the origin of the observed GZK cutoff of the cosmic ray proton spectrum \cite{Greisen:1966jv, Zatsepin:1966jv}.  
The \n\ production is dominated by the $\Delta^{+}$ resonance, which for CMB photons is realized at $E_p \gtrsim 5\times 10^{10}$ GeV of proton energy. Through the resonance,   pions are produced, and their decay chain generates muon and electron \ns.   Since the parent protons are absorbed efficiently, we expect that the \n\ flux can be higher than the observed proton one. The cosmogenic neutrino spectrum $\phi(E')$ has been calculated in Refs.~\cite{Engel:2001hd,Kalashev:2002kx}.

The neutrino emissivity for cosmogenic neutrinos is given by
\be\label{cosmogenic}
\mathcal{L}_{\nu}^{\rm cosm} = \mathcal{N}_{0} (1+z)^{n-1} \phi(E') 
\ee
where $\mathcal{N}_{0}$ and $n$ characterize the source population in normalization and redshift evolution. The \n\ spectrum $\phi(E')$ has an exponential cutoff at the maximum proton acceleration energy $E_{\rm max}$. 

Under the assumption that UHE protons are produced by stellar or galactic-size objects, the evolution of the source should have $n \simeq 3-4$, with a maximum redshift $z_{\rm max} \simeq 7-10$. Here we take $n=3$ and $z_{\rm max } = 10$, and use the single source spectrum from Ref.~\cite{Engel:2001hd}, which has maximum acceleration energy $E_{\rm max} \sim 10^{11}$ GeV.   The resulting diffuse flux is shown in Fig.~\ref{fluxeslimits}, and is practically the same with and without resonant absorption.  Neutrino-\n\ scattering effects are completely negligible, since the sources are at low redshifts, $z_{\rm max}\lta 10$, where the optical depth is very small, $\tau \ll 1$.  Besides,  even a modest \abs\ dip would probably be unobservable because the flux declines sharply with energy above $E_{\rm max}$, and is greatly suppressed in the part of the spectrum relevant for absorption, $E \sim 10^{12} - 10^{13} $ GeV. 

%============================================

\subsection{Gamma ray bursts and active galactic nuclei}

Gamma ray bursts (GRB) and Active Galactic Nuclei (AGN) are sources of high energy gamma rays,  and candidate sources of UHE neutrinos. The UHE neutrinos are produced via hadronic cascades in the interactions of high energy protons with the intense photon background in the source. The redshift evolution of these sources is believed to be stronger than the star formation rate history. Specifically, their comoving  rate can be written as:
\begin{equation}
\frac{dN}{dz}=A\cdot \eta_{SFR}(z)(1+z)^{\beta} ~\frac{dV_c}{dz}\frac{1}{1+z},
\label{grbrate}
\end{equation}
 where  $V_c$ is the comoving volume,  A is a normalization constant, and $\beta \simeq 1.5$ for GRB \cite{Robertson:2011yu} and $\beta \simeq 2$ for AGN \cite{Hasinger:2005sb}.  Here $\eta_{SFR} $
 is  star formation rate density \cite{Yuksel:2008cu}:
\begin{equation}
\eta_{SFR}(z)=\eta_0 \left[(1+z)^a+\left( \frac{1+z}{B} \right)^b+\left( \frac{1+z}{C} \right)^c \right]^{-0.1}~. 
\end{equation}
with $\eta_0=0.02~ M_{\odot}~ {\rm yr^{-1} Mpc^{-3}}$ ($M_{\odot}$ is the mass of the Sun), $a=-34,~ b=3,~ c=3.5,~ B=5000, ~C=9$ \cite{Yuksel:2008cu}. 

The neutrino spectra follows a power law, $\phi(E') \propto E'^{-2}$, with a lower and upper energy cutoffs. Therefore  the neutrino emissivity is given by,
\begin{equation}
\mathcal{L}_{\nu}^{\rm GRB} = {j}_{0} \frac{dN}{dz} \left(\frac{E (1+z)}{E_{\rm max}}\right)^{-2} \Theta[E(1+z)- E_{\rm min}] ~\Theta[E_{\rm max} - E(1+z)] ~. 
\end{equation}
Here we use the normalization $j_0\simeq  10^{-49}$GeV cm/s,   $E_{\rm min} \simeq 10^9$GeV, and $E_{\rm max}\simeq 10^{12}$GeV \cite{Eberle:2004ua}.  

Similar to cosmogenic \ns, our results show that \abs\ is negligible for GRB and AGN \ns, since their flux is dominated by small redshifts of order a few, and is cut off below the energy range of interest for absorption.

%========================================================================================
\section{Discussion}
\label{sec:disc}

We have discussed the \abs\ of UHE \ns\ due to scattering on the \cnb, with a focus on the effect of the thermal distribution of the background \ns.  The thermal effects have been fully modeled, using realistic \n\ mass spectra and mixings.  Neutrino \abs\ is dominated by the resonant $Z^0$ production. Thermal effects cause the resonance to be realized for an interval of the beam \n\ energy, depending on the scattering angle and the temperature of the background.  As a consequence, the suppression of the UHE \n\ spectrum changes from sharp to wide as the thermal effects become important.  This transition occurs when   $\bar p(z) \sim m_{\rm min}$, with $m_{\rm min}$ being the  smallest of the three \n\ masses.  In terms of cosmic time, this corresponds to redshift $1+ z_{\rm th} \sim 16~ m_{\rm min}/(10^{-2}~{\rm eV})$.  For $m_{\rm min} \lta 10^{-4}~{\rm eV}$, thermal effects are already important, for the lightest \n\ species, at the present time.  However, this does not translate in a flux suppression, due the insufficient optical depth. We find that the optical depth is substantial, $\tau \gta 1$, for \n\ sources at $z \gta 10$ [Eq.~(\ref{taunr})].

The fact that  $z \gta 10$ is required to have significant suppression has two important consequences. First, \ns\ from stellar and galactic sources (e.g., cosmogenic \ns\ and \ns\ from AGN and GRBs), which extend up to $z \sim 5$ or so, have negligible absorption, and therefore their spectrum is a  direct representation of the physics of the sources.  Secondly, an observable spectrum distortion should have at most two dips, not three.  This is because, at $z \gta 10$,  the mass difference between $m_1$ and $m_2$ is comparable with the average \n\ energy, i.e., $m_2 - m_1 \sim 10^{-2}~{\rm eV} \simeq \bar p $, therefore the scattering off $\nu_1$ and $\nu_2$ causes a single dip instead of two separate ones. 

A further smearing of the suppression dips is produced by integrating over the spatial distribution of the sources.  We worked out specific examples of diffuse UHE \n\ fluxes, with a focus on \ns\ from top down mechanisms, for which the sources extend beyond $z \sim 10$, and therefore a strong \abs\ is expected.  The cases considered were cosmic string kinks and cusps, super-heavy dark matter, cosmic necklaces and superconducting strings.    In all these models the flux is dominated by the contribution of sources closest to us, i.e., at the lowest redshift, $z_{\rm min}$, which, in general, depends on energy. Therefore, in first approximation  the flux suppression is described by $P(E, z_{\rm min})$, with $P(E, z)$ being the probability of transmission for a \n\ of energy $E$ (at Earth) and production epoch $z$ [Eq.~(\ref{pave3})].
We have found that, indeed, for sources with $z_{\rm min} \gta 10$, the diffuse flux is suppressed strongly, up to an order of magnitude or even more, in some cases. A broad  suppression  valley is localized between $10^{12}$ and $10^{14}$ GeV; its shape and extent in energy depends on the details of the model and on the \n\ mass spectrum. However,  the dependence on the \n\ mass spectrum, and especially on the mass hierarchy, is relatively weak. 
 This generality is a result of the thermal effects, which, at least for the hierarchical mass spectra, dominate over the \n\ mass effect, and tend to make the suppression mass-independent.  This has an immediate implication: the energy interval $10^{12} - 10^{13}$ GeV is potentially the worst place to look to discover UHE \ns!  This might have to be taken into account in the design of UHE \n\ detectors.  We note that  SKA (which is not optimized for \n\ detection) has maximum sensitivity exactly in this range (see Fig.~\ref{fluxeslimits}), therefore it might find itself in a position of disadvantage compared to other probes with different energy sensitivity.

Without being too specific, here we assume that UHE \n\ detectors can identify, at least roughly, a suppression in the \n\ spectrum.  In the worst case of energy-blind detectors, some sensitivity can be gained by comparing the fluxes measurements or upper limits from  different techniques probing different parts of the \n\ spectrum. For a single detector, a suppression may be defined only relative to a model of reference.  

If UHE \ns\ are detected, and the data are compatible with a suppression due to \n\ \abs, what can be learned from them?  Considering that the suppression bears only little dependence on the \n\ mass and mixing pattern, the main information will be on the physics of the sources. In particular, the observation of \n\ absorption will indicate, beyond doubt, a population of sources extending to $z \gta 10$, earlier than the time of formation of stars and galaxies. Therefore, this might be a way to discover, or further substantiate, the existence of cosmological relics like superheavy dark matter, cosmic strings or necklaces.   The detailed shape of the suppression dip (if available) would in principle allow to reconstruct $z_{\rm min}$ as a function of energy since the distortion is roughly determined by $P(E,z_{\rm min})$. This can help to discriminate between different source models, if combined with other elements like the presence of a minimum energy cutoff (favoring cosmic string cusps and kinks) or a high energy flux termination (which would favor cosmic necklaces and superheavy dark matter).  

Spectral distortions due to resonant absorption are, at least in principle, an interesting probe of the \cnb\ at relatively recent cosmological times, $z \sim 10-100$, that are out of the reach of both cosmological surveys [like those of Large Scale Structure ($z \lesssim 10$), and of the CMB ($z \sim 1100$), etc.] and  direct detections of the \cnb\ (e.g., by zero-threshold nuclear decay \cite{weinberg}, testing $z =0$). In particular, an observed absorption pattern could help to constrain, or even reveal, several exotic effects:\\
(i) {\it Non standard \n\ number density}. An increased population of active \ns\ would result in stronger absorption dips. For example, we could consider an increase in number density by a factor $4/3$, corresponding to an effective number of cosmological relativistic degrees of freedom $N_{\rm eff}=4$, which has recently attracted some interest (see e.g., \cite{Hinshaw:2012aka,Ade:2013zuv}). This increase would change the optical depth  by the same amount, and shorten the \n\ horizon down to $z \sim 120$.  A depletion of the \n\ population at late times is also possible, for example due to \n\ decay into a sterile \n\ or very weakly interacting species (e.g., \cite{Beacom:2004yd}). This would suppress the absorption and extend the \n\ horizon. \\
(ii) {\it Non-standard \n\ spectrum}.  Currently, there is no direct information on the \cnb\ spectrum, and indirect constraints are limited. Deviations from a thermal spectrum have been suggested, e.g., as a consequence of active-sterile \n\ conversion (e.g., \cite{Kishimoto:2006zk}). 
They would influence the shape of the absorption dips, which could be narrower for a narrower \n\ spectrum or if the spectrum is much colder than expected, so to make most of the \ns\ non-relativistic at the epochs of interest.\\
(iii) {\it Neutrino asymmetry, anomalous flavor composition, non-standard \n\ interactions, and other exotica.}  Our results could be generalized to consider a broader range of situations, including a \n\ population which is not flavor and CP-symmetric. Although these possibilities are interesting, to study them with  UHE \n\  absorption may be complicated by degeneracies between the physics of the \cnb\  and the physics of the sources: for example, there is a degeneracy between the \n\ number density and the redshift distribution of the sources such that they both affect the depth of the spectral dips in a similar way. 

If nothing else, it is important to accurately model the \abs\ dips to correctly interpret observations, and in particular to distinguish the effect of resonant $\nu - \bar \nu$ annihilation from spectral features of different nature, e.g., due to the overlap of two fluxes of different origin (bimodal spectrum), that could roughly mimic an absorption dip.  

 Although some of the effects described here require high precision and statistics, we can not underestimate the potential of this field to open a completely new way to explore the sky and learn about neutrinos.

%========================================================================================
\appendix
%=================
\section{Cross sections}
\label{app:cs}

\subsection{Resonant cross section}
The resonant neutrino-antineutrino annihilation ($\nu \bar{\nu} \rightarrow Z^0  \rightarrow f \bar f$) occurs in the s-channel. The cross section is expressed as a function of the Mandelstam variable, $s = (q^{\mu} + p^{\mu})^2$. Here $q^{\mu} = [E',~ {\bf {q}}]$ and $p^{\mu} = [\sqrt{{\bf {p}}^2 + m_j^2},~ {\bf {p}}]$ are the four momenta of an UHE \n\ and background neutrinos, respectively. Since for UHE neutrinos $\left|\bf {q} \right| \gg m_{\nu_{j}}$, $E' \approx \left|\bf {q} \right| \equiv q$, then its four momentum is simply $q^{\mu} =E' [1, {\bf \hat{q}}]$. Note that in an expanding universe, we replace $E' = E (1+z)$ and $p = p_0 (1+z)$, where $E$ and $p_0$ are the values of the beam energy and background \n\ momentum at present epoch, respectively. Then, the Mandelstam variable, $s$, in the comoving frame is: 
\be\label{s}
s(E',p,\theta) \approx 2 E' \left[ \sqrt{p^2 + m_j^2} - p \cos\theta \right],
\ee
where ${\bf \hat q} \cdot {\bf p} \equiv p \cos\theta$. The differential cross section for the resonant s-channel is \cite{D'Olivo:2005uh}
\be
d\sigma_{\rm r}(E',p,s)= \frac{G_F\Gamma M_Z}{\sqrt{2} E'\sqrt{p^2+m_{\nu_{j}}^2}} \frac{s (s-2m_{\nu_{j}} ^2)}{(s-M_{Z}^2)^2+\xi s^2}ds,
\label{diffcross}
\ee
where $G_{F} = 1.16637 \times 10^{-5} ~{\rm GeV}^{-2}$ is the Fermi coupling constant, $M_Z=91.1876$ GeV, $\Gamma=2.495$ GeV is the width of the $Z^0$ resonance, $\xi =\Gamma^2/M_Z^2$.  The total resonant cross section is obtained by integrating over s:
\be\label{sigmaresint}
\sigma_{\rm r}(E',p)= \int_{s_-}^{s_+} d\sigma_{\rm r}(E',p,s),
\ee
where $s_\pm \equiv 2 E' \left[ \sqrt{p^2 + m_j^2} \pm p \right]$ corresponding to head-on and parallel scattering, respectively. Eq.~(\ref{sigmaresint}) can be expressed in an analytical form \cite{D'Olivo:2005uh} 
\ba\label{sigmares}
\sigma_{\rm r}(E',p) &=& \frac{G_F \Gamma M_Z}{\sqrt{2}E'\sqrt{p^2+m_{\nu_{j}}^2}} \biggl [ \frac{s}{1+\xi}
- \frac{M_Z^2(\xi-1)}{\sqrt{\xi}(1+\xi)^2}~ {\rm arctan}\left( \frac{(1+\xi)~s -M_Z^2}{M_Z^2\sqrt{\xi}} \right) \nonumber\\ 
	&+& \frac{M_Z^2}{(1+\xi)^2}~ {\rm ln} \left[(1+\xi)s^2 -2 M_Z^2+ M_Z^4 s\right] \biggr] \bigg \vert_{s_-}^{s_+}~,
\ea
where we take $s - 2m_{\nu_{f}} ^2 \approx s$.
The resonant cross section $\sigma_{\rm r} (E',p)$ includes all kinematically allowed final states ($\bar q q, {\bar l} l$), which is taken into account in the width $\Gamma$.

%===============================
\subsection{Non-resonant cross sections}

Non-resonant cross sections are smooth functions of the beam energy $E'$, thus it is a very good approximation to use the value of $s$, averaged over scattering angle and momenta of the background \ns, instead of Eq.~(\ref{s}), to simplify the analysis: 
\be\label{aves}
\bar s(E',m_j) \equiv 2E'  \sqrt{\bar p^2 + m_j^2}.
\ee
All the relevant non-resonant processes are summarized as follows \cite{Roulet:1992pz,Barenboim:2004di}: The t-channel $Z$-exchange ($\nu_{\alpha}\bar{\nu}_{\beta} \to \nu_{\alpha}\bar{\nu}_{\beta}$) cross section with multiplicity $3$ (including $3$ different flavors for the target \n) is:
\be\label{sigmatz}
\sigma_{tZ}=3 \frac{G_{F}^2\bar s(E',m_j)}{2\pi}F_1(y_Z),
\ee
where $ F_1(y)=[y^2+2y-2(1+y)~ {\rm ln}(1+y)]/y^3$ and $y_Z = \bar s(E',m_j)/M_{Z}^2$. For $\alpha = \beta$, there is an s-t interference term with multiplicity $1$:
\be
\sigma_{stZ}= \frac{G_{F}^2\bar s(E',m_j)}{4\pi}F_2(y_Z)\frac{y_Z-1}{(y_Z-1)^2+\Gamma^2/M_{Z}^2}~,
\ee
where $F_2(y)=[3y^2+2y-2(1+y)^2~{\rm ln}(1+y)]/y^3$. The t-channel $W$-exchange ($\nu_{\alpha}\bar{\nu}_{\beta} \to l_{\alpha}\bar{l}_{\beta}$) cross section with multiplicity $3$ is:
\be\label{sigmatw}
\sigma_{tW}=3 \frac{2G_{F}^2\bar s(E',m_j)}{\pi}F_1(y_W),
\ee
where $y_W = \bar s(E',m_j)/M_{W}^2$ and $M_W = 80.385$ GeV. For $\alpha = \beta$, there is an interference between the s-channel $Z$-exchange and the t-channel $W$-exchange is with multiplicity 1:
\be
\sigma_{stZW}= \frac{2G_{F}^2({\rm sin}^2\theta_W -1/2)}{\pi}y_WM_{W}^2F_2(y_W)\frac{y_Z-1}{(y_Z-1)^2+\Gamma^2/M_{Z}^2}~,
\ee
where ${\rm sin}^2\theta_W=0.23149$. The elastic t-channel $Z$-exchange ($\nu_{\alpha} \nu_{\beta} \to \nu_{\alpha} \nu_{\beta}$) cross section with multiplicity $3$ is:
\be
\sigma_{tZ}^{\rm el}=3 \frac{2G_{F}^2M_{Z}^2}{2\pi}\frac{y_Z}{1+y_Z}~.
\ee
There is also the u-channel $Z$-exchange that contributes to the same process ($\nu_{\alpha} \nu_{\beta} \to \nu_{\alpha} \nu_{\beta}$) with multiplicity $1$:
\be
\sigma_{uZ}= \frac{2G_{F}^2\bar s(E',m_j)}{\pi}\left[\frac{1}{1+y_Z}+\frac{{\rm ln}(1+y_Z)}{y_Z (1+y_Z/2)}\right].
\ee
The weak charged vector boson pair production cross section in the s-channel $Z$-exchange and the t-channel $l$-exchange ($\nu_{\alpha}\bar{\nu}_{\alpha} \to W^+W^-$) with multiplicity $1$ and threshold $s > 4 M_W^2$ is:
\ba
\sigma_{WW} &=&  \frac{G_{F}^2y_WM_{W}^2\beta_W}{12\pi} \biggl [ \frac{\beta_{W}^2M_{W}^4}{M_{Z}^4(y_Z-1)^2}(12+20y_W+y_{W}^2)\\ \nonumber
&+& \frac{2M_{W}^2}{M_{Z}^2(y_Z-1)y_{W}^2} \left (24+28y_W-18y_{W}^2-y_{W}^3 +\frac{48(1+2y_W)L_W}{\beta_Wy_W} \right)\\ \nonumber
&+&\frac{1}{y_{W}^2} \left (y_{W}^2+20y_W-48-\frac{48(2-y_W)L_W}{\beta_Wy_W} \right) \biggr ],
\ea
where $\beta_W=\sqrt{1-4/y_W}$ and $L_W= {\rm ln} [(1+\beta_W) (1-\beta_W)]$. The weak neutral vector boson pair production cross section in the s-channel ($\nu_{\alpha}\bar{\nu}_{\alpha} \to ZZ$) with multiplicity 1 and threshold $s > 4 M_Z^2$ is:  
\be
\sigma_{ZZ}=  \frac{G_{F}^2M_{Z}^2}{\pi}  \frac{\beta_Z}{y_Z-  2} \left(\frac{2}{y_Z}-1+\frac{1+y_Z^2}{2y_Z^2\beta_Z}L_Z \right),
\ee
where $\beta_Z=\sqrt{1-4/y_Z}$ and $L_Z= {\rm ln} [(1+\beta_Z) (1-\beta_Z)]$. Finally, $ZH$ production cross section in the s-channel ($\nu_{\alpha}\bar{\nu}_{\alpha} \to Z^0H $) with multiplicity 1 and threshold $s >  (M_Z+M_H)^2$ is:
\be
\sigma_{ZH}=  \frac{G_{F}^2M_{Z}^2}{96\pi}  \frac{\sqrt{\lambda} \beta_Z}{y_Z} \frac{\lambda y_Z + 12}{(y_Z - 1)^2}~,
\ee
where $\lambda = [1-(M_{H}+M_{Z})^2 / (y_Z M_{Z}^2)] [1- (M_{H}-M_{Z})^2 / (y_Z M_{Z}^2)]$ and $M_H = 125$ GeV. The total cross section is  the sum of all the resonant and non-resonant channels, as shown in Fig.~\ref{xsectionpartial}. 

%=====================
\section{Scattering amplitude and rate}
\label{app:amplitude}

In this appendix, we schematically show the dependence of the scattering rate on the \n\ mixing matrix for an UHE neutrino in a flavor eigenstate $\nu_{\alpha}$ and a \cnb\ neutrino in a mass eigenstate $\nu_{j}$, given by Eq.~(\ref{GammaTot}). For simplicity, consider the scattering amplitude for the s-channel process, $\nu_{\alpha} \bar \nu_{j} \to f \bar f$, where $f$ is a final state fermion. The scattering amplitude $M_{\alpha j}$ is proportional to
\be
M_{\alpha j} \propto \bra{\bar \nu_{j}} O \ket{\nu_{\alpha}} = \sum\limits_{i} U^\ast_{\alpha i}~ e^{-i \Phi_i(t)}  \bra{\bar \nu_{j}} O \ket{\nu_{i}}  ~,
\label{malpha}
\ee
where $U_{\alpha i}^*$ are the elements of the neutrino mixing matrix and $\Phi_i(t)=\int^{t}_{t_i} dt'~ \sqrt{[p(t')]^2 + m_i^2}$ is the quantum phase due to the \n\ propagation in vacuum between the time of production, $t_i$, and the time  $t$ when the collision occurs.  This phase is responsible for \n\ flavor oscillations.  

In Eq.~(\ref{malpha}), the nonvanishing elements are the diagonal ones, i.e., $\bra{\bar \nu_{i}} O \ket{\nu_{j}} \propto \delta_{ij} M_j$. Hence, 
\be
M_{\alpha j} \propto U^\ast_{\alpha j} ~ e^{-i \Phi_i(t)}  \bra{\bar \nu_{j}} O \ket{\nu_{j}} \propto  U^\ast_{\alpha j} ~ e^{-i \Phi_i(t)}  M_j~. 
\ee 
The corresponding cross section is then 
$\sigma(m_j) \propto |M_j|^2$. Note that  the phase $\Phi_j$ cancels, hence neutrino oscillations do not affect the cross section provided that the background neutrino is in mass eigenstate as we discussed in Sec.~\ref{sec:cnb}. Then, given $dn$ as the number density of the \cnb\ \ns\ of each species, we have the scattering rate for an UHE neutrino of flavor $\alpha$  in the \cnb\ for a given process:  
\ba
d\Gamma_\alpha =dn \sum\limits_{j=1}^3  |U_{\alpha j}|^2   \sigma(m_j), 
\ea
which, after integration over the \n\ spectrum, recovers Eq.~(\ref{GammaTot}).

%========================================================================================
\acknowledgments
We would like to thank Ken Olum, Hiroyuki Tashiro and Tanmay Vachaspati for useful discussions. This work is supported in part by the National Science Foundation Grants No.~PHY-0854827 and No.~PHY-1205745 and Department of Energy at Arizona State University.

%=========================================================================================

\end{document}